\documentclass[10pt,a4paper,preprint,english,aip,jcp,preprint,superscriptaddress,onecolumn]{revtex4-1}
\usepackage[T1]{fontenc}
\usepackage[latin9]{inputenc}
\usepackage{color}
\usepackage{verbatim}
\usepackage{amsmath}
\usepackage{graphicx}
\usepackage{amssymb}
\usepackage{esint}

\makeatletter

 \@ifundefined{textcolor}{}
 {%
   \definecolor{BLACK}{gray}{0}
   \definecolor{WHITE}{gray}{1}
   \definecolor{RED}{rgb}{1,0,0}
   \definecolor{GREEN}{rgb}{0,1,0}
   \definecolor{BLUE}{rgb}{0,0,1}
   \definecolor{CYAN}{cmyk}{1,0,0,0}
   \definecolor{MAGENTA}{cmyk}{0,1,0,0}
   \definecolor{YELLOW}{cmyk}{0,0,1,0}
 }

\@ifundefined{definecolor}
 {\usepackage{color}}{}

\newcommand{\eg}{e.g.}
\newcommand{\ie}{i.e.}

\newcommand{\vs}{vs.}

\newcommand{\etal}{\textit{et al.}}
\newcommand{\bvec}[1]{\mathbf{#1}}
\newcommand{\vlocr}{V_{\mathrm{locps}}\left(\bvec{r}\right)}

\newcommand{\vlocIr}{V_{\mathrm{locps},I}\left(\bvec{r}\right)}
\newcommand{\vlocIx}{V_{\mathrm{locps},I}\left(\bvec{x}\right)}
\newcommand{\vlocIxscalar}{V_{\mathrm{locps},I}\left(x\right)}
\newcommand{\vlocsxscalar}{V_{\mathrm{locps},s}\left(x\right)}
\newcommand{\vlocIg}{\tilde{V}_{\mathrm{locps},I}\left(\bvec{G}\right)}
\newcommand{\vlocIgscalar}{\tilde{V}_{\mathrm{locps},I}\left(G\right)}
\newcommand{\vlocsgscalar}{\tilde{V}_{\mathrm{locps},s}\left(G\right)}

\newcommand{\gcut}{G_{\mathrm{cut}}}
\newcommand{\cell}{\Omega}

\makeatother

\usepackage{babel}

\begin{document}

\title{\textcolor{blue}{Electrostatic Interactions in Finite Systems treated
with Periodic Boundary Conditions: Application to Linear-Scaling Density
Functional Theory}}

\author{N. D. M. Hine}

\affiliation{Department of Physics and Department of Materials, Imperial College
London, Exhibition Road, London SW7 2AZ, UK.}

\author{J. Dziedzic}

\affiliation{School of Chemistry, University of Southampton, Highfield, Southampton
SO17 1BJ, UK.}

\altaffiliation{[Also at] Faculty of Technical Physics and Applied Mathematics, Gdansk University of Technology, Poland.}

\author{P. D. Haynes}

\affiliation{Department of Physics and Department of Materials, Imperial College
London, Exhibition Road, London SW7 2AZ, UK.}

\author{C. K. Skylaris}

\affiliation{School of Chemistry, University of Southampton, Highfield, Southampton
SO17 1BJ, UK.}

\date{\today}
\begin{abstract}
We present a comparison of methods for treating the electrostatic
interactions of finite, isolated systems within periodic boundary
conditions (PBCs), within Density Functional Theory (DFT), with particular
emphasis on linear-scaling (LS) DFT. Often, PBCs are not physically
realistic but are an unavoidable consequence of the choice of basis
set and the efficacy of using Fourier transforms to compute the Hartree
potential. In such cases the effects of PBCs on the calculations need
to be avoided, so that the results obtained represent the open rather
than the periodic boundary. The very large systems encountered in
LS-DFT make the demands of the supercell approximation for isolated
systems more difficult to manage, and we show cases where the open
boundary (infinite cell) result cannot be obtained from extrapolation
of calculations from periodic cells of increasing size. We discuss,
implement and test three very different approaches for overcoming
or circumventing the effects of PBCs: truncation of the Coulomb interaction
combined with padding of the simulation cell, approaches based on
the minimum image convention, and the explicit use of Open Boundary
Conditions (OBCs). We have implemented these approaches in the ONETEP
LS-DFT program and applied them to a range of systems, including a
polar nanorod and a protein. We compare their accuracy, complexity,
and rate of convergence with simulation cell size. We demonstrate
that corrective approaches within PBCs can acheive the OBC result
more efficiently and accurately than pure OBC approaches.
\end{abstract}
\maketitle

\section{Introduction}

Density Functional Theory (DFT) \cite{hohenberg_inhomogeneous_1964,kohn_self-consistent_1965}
is widely and routinely used for computational electronic structure
simulations due to its favorable balance of speed and accuracy. However,
making DFT simulations scale well to the numbers of atoms required
to study large complex systems such as proteins and nanostructures
presents significant challenges. Various linear-scaling approaches
to DFT have emerged over the last two decades to meet this challenge
\cite{goedecker_linear_1999,yang_direct_1991,galli_large_1992,li_density-matrix_1993,ordejon_linear_1995,hernandez_self-consistent_1995,fattebert_towards_2000,skylaris_non-orthogonal_2002}.
Several of these methods use basis sets which are related to plane
waves and require periodic boundary conditions (PBCs). The plane-wave
pseudopotential approach has been developed with crystalline systems
in mind, and as these are genuinely periodic, the treatment of electrostatics
in the framework of PBCs was a natural choice with significant advantages.
In reciprocal space, the Hartree interaction is diagonal, so the Hartree
potential and energy are easily obtained using Fast Fourier Transforms
(FFTs). Furthermore, the plane-wave basis set is systematic in the
sense that it provides a uniform description of space and can be improved
by increasing the value of one parameter.

However, the increasing use of linear-scaling DFT (LS-DFT) in large
systems highlights long-standing issues in electronic structure methods
relating to the treatment of electrostatic interactions, \ie~the
long-ranged parts of the Coulomb interaction between electron density
and electron density (`Hartree' terms), electron density and ion cores,
and between ion cores, under PBCs.

Bulk systems can be genuinely periodic and then the influence of periodic
replicas is desired; however, to allow simulation of finite, isolated
systems within PBCs, the supercell approximation is widely used \cite{cohen_self-consistent_1975,ihm_momentum-space_1979,payne_iterative_1992}.
This involves the replacement of a genuinely isolated system with
a lattice of periodic replicas, with vacuum `padding' surrounding
the system to reduce the influence of periodic replicas on each other.
While this is a reasonable approach, it introduces finite size errors
whereby the total energy varies with supercell size.

The use of a supercell is frequently a well-controlled approximation:
that is to say, by increasing the size of the cell and thus the distance
between periodic images, one rapidly approaches the true isolated,
non-periodic limit. For example, in the case of relatively small,
charge-neutral molecules without significant dipole moment, one needs
to ensure simply that the charge densities of periodic replicas do
not overlap to any significant extent. In other cases, the amount
of vacuum padding required to reach this limit can become prohibitively
large. The slow decay of the interaction of periodic replicas of a
monopole charge, as $1/R$, means that the infinite limit is impossible
to reach in practice for charged systems. Similarly, for highly-elongated
charge-neutral systems possessing a large dipole moment (such as in
a simulation of a polar semiconductor nanorod), the simulation cell
would likewise need to be unfeasibly large to prevent interactions
between periodic images of adjacent rods. Clearly the isolated limit
cannot always be found simply by extrapolating to infinite supercell
size. This issue is exacerbated as the isolated molecules and their
dipole moments become larger.

To address this problem, a large range of techniques that aim to either
reduce or eliminate the effects of the PBCs on the electrostatics
of grid-based electronic structure calculations have been developed
over the recent years \cite{hockney_computer_1981,m._leslie_energy_1985,makov_periodic_1995,jarvis_supercell_1997,kantorovich_elimination_1999,bengtsson_dipole_1999,martyna_reciprocal_1999,nozaki_energy_2000,schultz_charged_2000,a_castro_a_rubio_m_j_stott_solution_2003,rozzi_exact_2006,genovese_efficient_2006,ismail-beigi_truncation_2006,wright_comparison_2006,genovese_efficient_2007,dabo_electrostatics_2008,yu_equivalence_2008}.
These include methods which attempt to formulate an \emph{a posteriori}
correction term to add to the energy \cite{m._leslie_energy_1985,makov_periodic_1995,kantorovich_elimination_1999}
on the basis of a multipole expansion of the localised charge, having
first inserted a uniform periodic background to counter any monopole
charge \cite{bar-yam_electronic_1984}; methods which formulate a
more complex form of `counter-charge' which counteracts the periodic
interactions \cite{schultz_charged_2000,nozaki_energy_2000,dabo_electrostatics_2008,bengtsson_dipole_1999,yu_equivalence_2008},
and methods that modify the form of the interaction in real or reciprocal
space in order to avoid the existence of periodic interactions in
the first place \cite{jarvis_supercell_1997,martyna_reciprocal_1999,a_castro_a_rubio_m_j_stott_solution_2003,rozzi_exact_2006,genovese_efficient_2006,genovese_efficient_2007}.
%
{}

In this paper, we examine, implement and compare three different approaches
fulfilling these criteria: truncation of the Coulomb interaction in
real space, referred to here as `Cutoff Coulomb' (CC) \cite{jarvis_supercell_1997,rozzi_exact_2006};
the approaches of Martyna and Tuckerman (MT) and Genovese \emph{et
al., }which replace the periodic Coulomb interaction with a Minimum
Image Convention (MIC) approach to the Coulomb potential \cite{martyna_reciprocal_1999};
and the replacement of PBCs with Open Boundary Conditions (OBCs) using
a multigrid approach to the Poisson equation\cite{beck_real-space_2000,Trottenberg-2001-631,Brandt-1977-333}.
These methods are implemented and tested on a range of systems representing
typical cases with challenging electrostatic properties. We compare
their accuracy, convergence properties, complexity and computational
overhead, and summarise the advantages and disadvantages of each.

Throughout this work, we employ linear-scaling DFT with the ONETEP
code \cite{skylaris_introducing_2005}, and while our findings will
be applicable to all electronic structure methods, linear-scaling
or otherwise, we focus in particular on the challenges encountered
applying these methods to large, complex systems. System size can
be measured either by the number of atoms $N$ included in the simulation,
or by the volume $V$ of the simulation cell --- the latter being
particularly relevant in the case of isolated systems. ONETEP combines
linear-scaling computational effort, in that the total computational
time for a simulation of $N$ atoms can be made to scale as $O(N)$,
with near-independence of the computational effort on the amount of
vacuum padding (\ie nearly independent of $V$ at fixed $N$), and
systematic control of the accuracy with respect to the basis, akin
to that of plane-wave DFT. The requirements on any method used to
treat electrostatic interactions are therefore that it must have systematically
controllable accuracy, must not impose too high a computational overhead,
and must have low-order scaling with $N$ and $V$.

\section{Electrostatics in Linear-Scaling Density Functional Theory\label{sec:LS-DFT}}

The calculations in this work are performed with the ONETEP Linear-Scaling
DFT approach. Like most linear-scaling approaches to DFT, \textsc{ONETEP}
uses the density matrix rather than eigenstates of the Hamiltonian,
representing the single-electron density matrix $\rho(\mathbf{r},\mathbf{r}')$
in terms of nonorthogonal localised orbitals $\phi_{\alpha}(\mathbf{r})$
and a `density kernel' $K^{\alpha\beta}$ as\begin{equation}
\rho(\mathbf{r},\mathbf{r}')=\phi_{\alpha}(\mathbf{r})K^{\alpha\beta}\phi_{\beta}(\mathbf{r}')\;.\label{eq:density_matrix}\end{equation}
The Einstein convention of summation over repeated Greek indices will
be employed throughout. Using the density matrix, the electron density
$n(\mathbf{r})$ can be found from\begin{equation}
n(\mathbf{r})=\rho(\mathbf{r},\mathbf{r})=\phi_{\alpha}(\mathbf{r})K^{\alpha\beta}\phi_{\beta}(\mathbf{r})\;.\label{eq:density}\end{equation}
Where ONETEP differs from most linear-scaling approaches is that the
local orbitals, referred to as Nonorthogonal Generalised Wannier Functions
(NGWFs) \cite{skylaris_nonorthogonal_2002}, are themselves expressed
in a systematic underyling basis of periodic-sinc functions (psincs),
and are therefore systematically convergeable. This is achieved by
a double-loop optimisation\cite{skylaris_implementation_2006} of
both the coefficients $C_{i\alpha}$ of the psinc functions $D_{i}(\mathbf{r})$
describing each NGWF and the elements of the density kernel $K^{\alpha\beta}$:\begin{equation}
E_{\mathrm{T}}=\min_{\{C_{i\alpha}\}}L(\{C_{i\alpha}\})\;,\label{eq:emin}\end{equation}
 where $L$ represents optimisation with respect to the density kernel,
a generalisation of the occupancies, through:\begin{equation}
L(\{C_{i\alpha}\})=\min_{\{K^{\alpha\beta}\}}E(\{K^{\alpha\beta}\};\{C_{i\alpha}\})\;.\label{eq:lmin}\end{equation}
This results in a method with controllable accuracy and systematic
convergence of total energies and forces with respect to basis size,
equivalent to the plane-wave approach \textbf{\cite{skylaris_achieving_2007,hine_accurate_2011}},
in systems of tens of thousands of atoms \textbf{\cite{hine_linear-scaling_2009,hine_linear-scaling_2010}}.
Convergence is controlled by varying the spacing of the psinc grid,
in a manner equivalent to varying a plane-wave cutoff, described by
a cutoff energy $E_{\text{cut}}$, and by varying the cutoff radii
of the spherically-truncated NGWFs, described by a sphere radius $R_{\phi}$.
To achieve true asymptotically linear scaling behaviour, it is also
necessary to truncate the range of the density kernel $K^{\alpha\beta}$
so that elements for NGWFs centred on distant atoms for which $|\mathbf{R}_{\alpha}-\mathbf{R}_{\beta}|>R_{\mathrm{K}}$
are set to zero. However, this latter form of truncation is only necessary
in very large systems and will not be considered in this work.

This accurate and systematic approach to linear-scaling total energy
calculations demands that all aspects of the calculation be carried
out with high accuracy, including the long-range electrostatic part.
The electrostatic energy comprises the Hartree term, $E_{\mathrm{H}}[n]$,
which is the classical density-density interaction; the local pseudopotential
term, $E_{\mathrm{locps}}[n]$, which is the interaction of the electron
density with the long-ranged part of the potential resulting from
the ion cores; and the interaction between the ion cores, $E_{\mathrm{ion-ion}}$.
It should be noted that during the optimisation of the kernel and
NGWF coefficients $K^{\alpha\beta}$ and $C_{i\alpha}$, the full
interacting energy is minimised by conjugate gradients process, meaning
that no mixing of densities is required at any point. The problem,
then, becomes one simply of evaluating $E_{\mathrm{H}}[n]$ and $V_{\mathrm{H}}[n](\mathbf{r})$
for a given density $n(\mathbf{r})$ (which always integrates fo the
number of electrons $N_{e}$).

To be absolutely clear on the formalism involved, we will briefly
re-visit the standard approach, making careful distinctions on how
the expressions and their meaning vary under PBCs and under OBCs,
where the potentials tend to zero at infinity. In both cases, the
Hartree energy can be obtained as $E_{\mathrm{H}}=\tfrac{1}{2}\int n(\mathbf{r})V_{\mathrm{H}}(\mathbf{r})\, d\mathbf{r}$,
where the Hartree potential $V_{\mathrm{H}}(\mathbf{r})$ resulting
from a density $n(\mathbf{r})$, is formally obtained by solving the
Poisson equation: \begin{equation}
\nabla^{2}V_{\mathrm{H}}(\mathbf{r})=-4\pi n(\mathbf{r})\;.\label{eq:Poisson}\end{equation}
Note that we are working in atomic units, for which $1/\varepsilon_{0}=4\pi$.
This can in general be solved through the use of the corresponding
Green function $G(\mathbf{r},\mathbf{r}')=-1/4\pi|\mathbf{r}-\mathbf{r}'|$,
producing\[
V_{\mathrm{H}}(\mathbf{r})=-\int_{\text{all space}}\frac{n(\mathbf{r}')}{|\mathbf{r}-\mathbf{r}'|}\mathrm{d}\mathbf{r}'\;.\]
This result builds in the OBC definition that the potential goes to
zero at infinity, and cannot be used directly to evaluate $E_{\mathrm{H}}$
or $V_{\mathrm{H}}(\mathbf{r})$ under PBCs as the integral has infinite
value at all $\mathbf{r}$ for periodic $n(\mathbf{r}')$.

When PBCs are used Eq.~(\ref{eq:Poisson}) is only valid for charge
distributions of zero charge per simulation cell. If the total charge
on one cell $q=\int_{{\cell}}n(\mathbf{r})\, d\mathbf{r}$ is non-zero,
Eq.~(\ref{eq:Poisson}) is modified to the following form: \begin{equation}
\nabla^{2}V_{\mathrm{H}}(\mathbf{r})=-4\pi(n(\mathbf{r})-q/\cell)\;,\label{eq:Poisson_2}\end{equation}
where $\cell$ is the volume of the simulation cell. This is equivalent
to the insertion of a uniform background charge density of equal and
opposite charge to $n(\mathbf{r})$ so that the total charge is zero.
A periodic density will result in a periodic potential and in this
case we can re-write both sides of Eq.~(\ref{eq:Poisson_2}) in terms
of their discrete Fourier transforms and rearrange to obtain \begin{equation}
\tilde{V}_{\mathrm{H}}(\mathbf{G})=\frac{4\pi}{\Omega G^{2}}\left(\tilde{n}(\mathbf{G})-q\delta_{\mathbf{G},0}\right)\;.\label{eq:HartreePeriodic}\end{equation}
Note that Eq.~(\ref{eq:HartreePeriodic}) makes clear the utility
of a reciprocal space approach to calculating $\tilde{V}_{\mathrm{H}}(\mathbf{G})$,
even outside of a genuinely periodic situation: the Coulomb interaction
is diagonal in reciprocal space, so $\tilde{V}_{\mathrm{H}}(\mathbf{G})$
can be obtained trivially from $\tilde{n}(\mathbf{G})$. After obtaining
$V_{\mathrm{H}}(\mathbf{r})$ by an inverse FFT, the integral $E_{\mathrm{H}}=\tfrac{1}{2}\int_{{\cell}}n(\mathbf{r})V_{\mathrm{H}}(\mathbf{r})\, d\mathbf{r}$
can be performed only over one simulation cell to obtain the Hartree
energy per simulation cell.

\begin{figure}
\centering\includegraphics[clip,width=1\columnwidth]{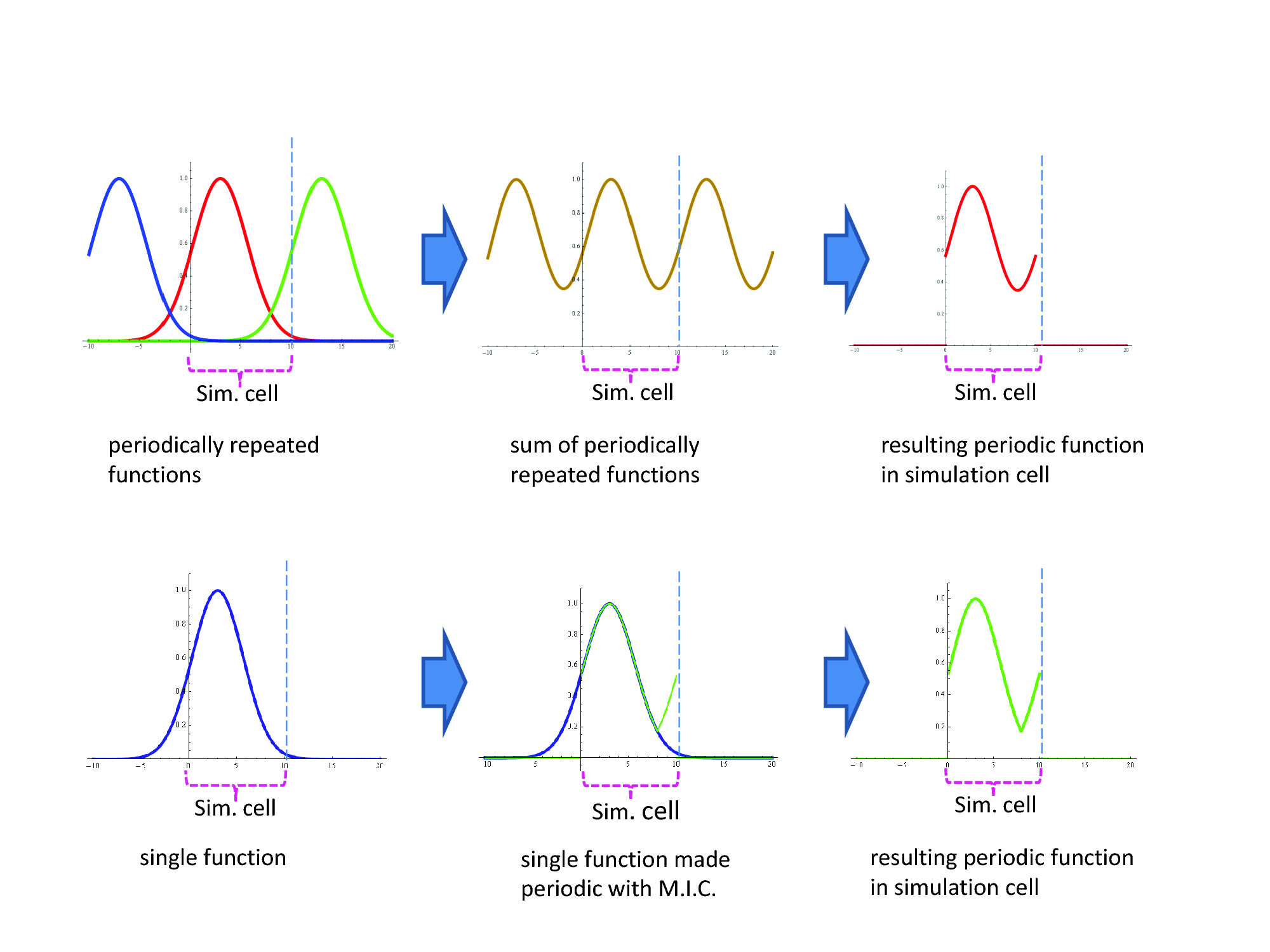}
\caption{\label{fig:ft-mt} Different ways of making a function obey periodic
boundary conditions inside a simulation cell, demonstrated for a Gaussian
function. Top panel: The Fourier transform approach. The resulting
function is the same as the one that would be obtained by a superposition
(sum) of periodically repeated Gaussians. Bottom panel: The Minimum
Image Convention (MIC) approach: the resulting function is the same
as the one that would be obtained by having a single Gaussian in the
simulation cell and making it periodic by applying the MIC.}

\end{figure}

In PBCs the potential is, by definition, the result of contributions
from not just the $n(\mathbf{r})$ of the home simulation cell but
also from the densities of an infinite number of periodic replicas
of that cell. A periodic function that can be constructed in this
way is demonstrated with the example at the top panel of Figure~\ref{fig:ft-mt}.
As we have already mentioned, the potential and the electrostatic
energy diverge for non-zero total charge in the simulation cell (or
equivalently when $\tilde{n}(\mathbf{G}=0)$ is nonzero). To avoid
this divergence one must set $\tilde{n}(\mathbf{G}=0)$ to zero for
each component making up total charge density (including the ion charges)
to ensure that the result is finite. Having made this choice however,
one alters the problem being studied as the potential $V_{\mathrm{H}}(\mathbf{r})$
obtained is that resulting not just from the infinite periodic array
of $n(\mathbf{r})$, but also from a neutralising charge distribution,
which is usually taken to be a uniform background charge over the
whole cell.

The same arguments apply to the other electrostatic terms, by replacing
the electron density $n(\mathbf{r})$ with the charge density of the
ions, in the form of a collection of point charges. For an isolated
system, the energy of interaction of the ions is of course simply
\begin{equation}
E_{\text{ion-ion}}=\frac{1}{2}\sum_{I,\, J\neq I}\frac{Z_{I}Z_{J}}{|\mathbf{R}_{I}-\mathbf{R}_{J}|}\;,\label{eq:eionion}\end{equation}
while under PBCs, in the presence of the neutralising background,
the energy of interaction per unit cell is most commonly calculated
using the Ewald technique \cite{allen_computer_1989}.

The influence of periodic neighbours will affect (polarise) the charge
distribution during a self-consistent electronic structure calculation.
Therefore, it should be immediately clear that no \emph{a posteriori}
approach to correcting total energies obtained from a simulation under
PBCs can be completely successful in providing total energies that
match those of an isolated system as even after the ``removal'' of
the periodicity the density will remain distorted to what it was in
the periodic calculation. Here we examine three approaches that are
applied within the self-consistent procedure and therefore are able
to correct not only the energy but also the potential.

\section{Cutoff Coulomb interactions\label{sec:Cutoff-Coulomb}}

One way to avoid the effects of PBCs which are intrinsic to the discrete
Fourier representation of the Coulomb potential is to use a modified
form for the Coulomb potential. One such possibility is the use of
a \textquotedbl{}cutoff\textquotedbl{} form of the Coulomb interaction.
This allows the usual Fourier transform-based approach to be used,
including a nominally periodic simulation cell, but truncates the
Coulomb potential so that it is confined within the primary simulation
cell. The approach has been applied by several previous works \cite{jarvis_supercell_1997,rozzi_exact_2006}
and is implemented in several codes \cite{marques_octopus:_2003,castro_octopus:_2006}.

The essence of the cutoff Coulomb approach is that the periodic, background-neutralised
Coulomb potential $V_{\mathrm{Ew}}(\mathbf{r})$ is replaced with
the bare Coulomb interaction, truncated so as to prevent any part
of the simulation cell feeling the potential from any neighbouring
copy. This removes the need for the canceling background, even though
the charge density is periodically repeated. Some new complications
arise however as the cutoff Coulomb potential needs to be generated
in reciprocal space.

To retain the simplicity of having an interaction that is diagonal
in reciprocal space, but still avoid the influence of periodic replicas,
one can use the following form for the Coulomb potential \begin{equation}
V_{\mathrm{CC}}(\mathbf{r}-\mathbf{r}')=\begin{cases}
\frac{1}{|\mathbf{r}-\mathbf{r}'|} & \qquad\mathbf{r}-\mathbf{r}'\in\mathcal{R}_{1}\\
0 & \qquad\mathbf{r}-\mathbf{r}'\notin\mathcal{R}_{1}\end{cases}\;.\label{eq:v_CC}\end{equation}
 $\mathcal{R}_{1}$ is a region of a size and shape chosen such that
when centered at any point $\mathbf{r}$ at which $V_{\mathrm{H}}(\mathbf{r})$
is required (this may be anywhere inside the main simulation cell,
or it may just be anywhere where the density is nonzero), $\mathcal{R}_{1}$
encloses all $\mathbf{r}'$ for which $n(\mathbf{r}+\mathbf{r}')\neq0$.
The Hartree potential is now obtained as the convolution of the cut-off
Coulomb operator and the density \begin{equation}
V_{\mathrm{H}}(\mathbf{r})=\int_{\cell}n(\mathbf{r}')V_{\mathrm{CC}}(\mathbf{r}-\mathbf{r}')\,\mathrm{d}\mathbf{r}'\;.\label{eq:vH_CC}\end{equation}
 The simplest shape for $\mathcal{R}_{1}$ is a sphere of radius $R_{\mathrm{c}}$,
for which $V_{\mathrm{CC}}^{\mathrm{sphere}}(\mathbf{r})=\Theta(|\mathbf{r}|-R_{\mathrm{c}})/|\mathbf{r}|$
where $\Theta$ is the Heaviside step function. In this case, the
Fourier transform of the interaction is well-known: \begin{equation}
\tilde{V}_{\mathrm{CC}}^{\mathrm{sphere}}(\mathbf{G})=\frac{4\pi(1-\cos(GR_{\mathrm{c}}))}{\Omega G^{2}}\;.\label{eq:v_CC_sphere}\end{equation}
 As this function does not have a singularity at $\mathbf{G}=0$ the
Hartree potential is obtained in reciprocal space as its product with
$\tilde{n}(\mathbf{G})$ as in Eq.~(\ref{eq:HartreePeriodic}) but
without the $q$ term as there is no longer the need to include a
uniform background charge. A spherical cutoff removes the periodicity
in all three spatial dimensions. If periodicity is retained in one
or two dimensions there are corresponding forms for $\tilde{V}_{\mathrm{CC}}(\mathbf{G})$
to account for these wire (1D periodicity) and slab (2D periodicity)
geometries. A comprehensive study was made by Rozzi \emph{et.~al.~}\cite{rozzi_exact_2006}
describing the terms of the cutoff Coulomb interaction for each geometry.

In a practical calculation, the electron density $n(\mathbf{r}$)
on a real space grid over the original simulation cell is transferred
to a grid for a larger `padded' cell of size $L_{\mathrm{pad}}$ and
padded with zeros, then Fourier transformed to give $\tilde{n}_{\mathrm{pad}}(\mathbf{G})$.
The terms of $\tilde{V}_{\mathrm{CC}}(\mathbf{G})$ are calculated
for this reciprocal space grid in advance and stored, and are used
to multiply the Fourier components $\tilde{n}_{\mathrm{pad}}(\mathbf{G})$
whenever the Hartree potential is required. Reverse Fourier transforming
these components gives $V_{\mathrm{H},\mathrm{pad}}(\mathbf{r})$
from which the values of $V_{\mathrm{H}}(\mathbf{r})$ on the original
cell are extracted.

The corresponding cut-off form of the Coulomb interaction must also
be used in place of the long-ranged Coulombic tail of the ion cores
in the local pseudopotential $V_{\mathrm{locps}}(\mathbf{r}).$ To
achieve this, $\tilde{V}_{\mathrm{locps}}(\mathbf{G})$ is calculated
over the whole padded grid, replacing the $\frac{4\pi}{\Omega G^{2}}Z_{\mathrm{ion}}$
term by $\tilde{V}_{\mathrm{CC}}(\mathbf{G})Z_{\mathrm{ion}}$ for
the relevant form of cutoff Coulomb interaction. This is then transformed
to real space by Fourier transform and extracted to the standard grid
to give the required form of $V_{\mathrm{locps}}(\mathbf{r})$. Similarly,
the periodic Coulomb and Ewald terms in the calculation of the forces
acting on the ion cores are replaced by their cutoff Coulomb forms.

\begin{figure}
\includegraphics[width=0.6\columnwidth]{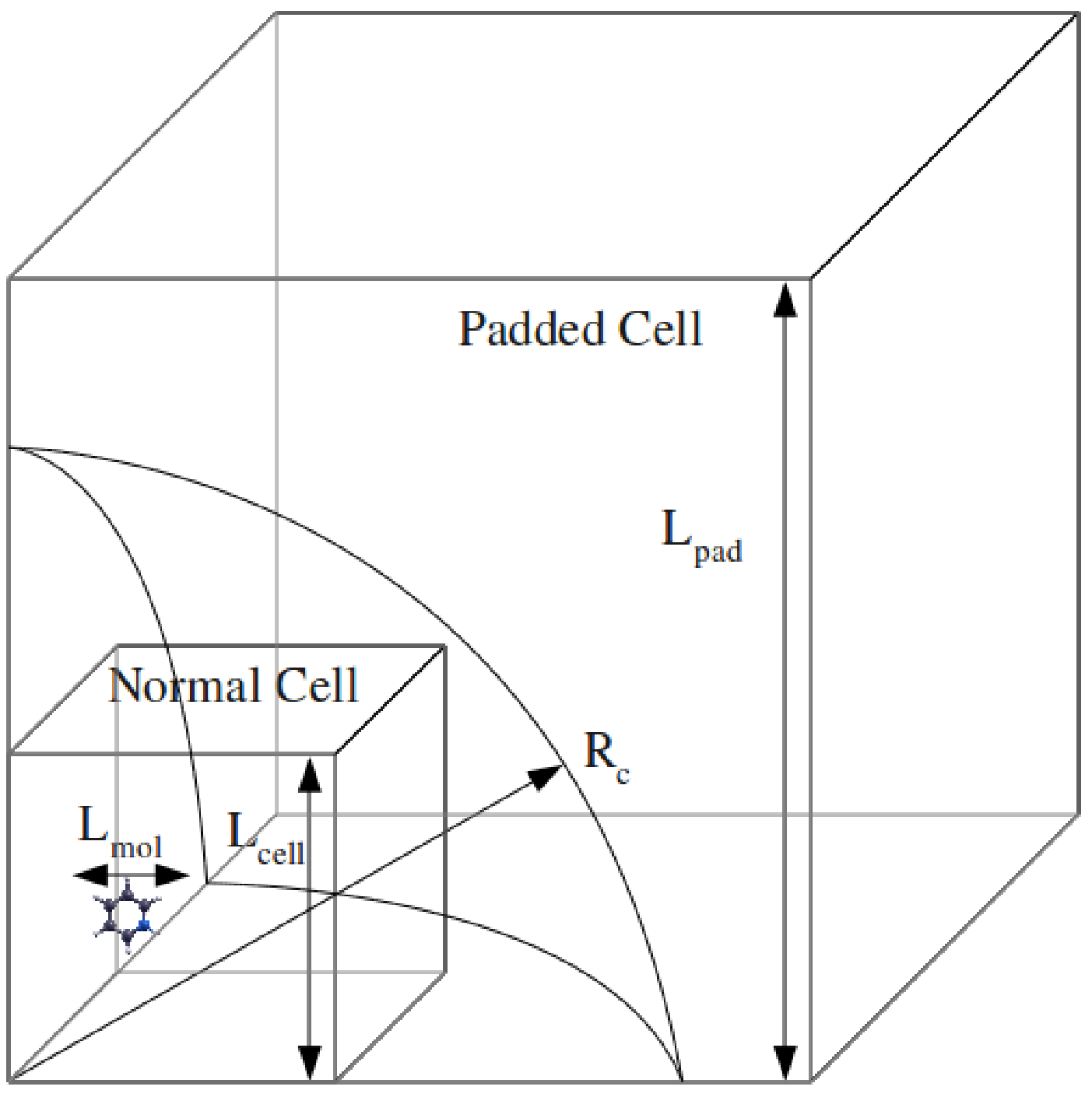}

\caption{Illustration of the cell sizes $L_{\mathrm{cell}}$, $L_{\mathrm{pad}}$
and cutoff radius $R_{\mathrm{c}}$ required for the spherical cutoff
Coulomb approach. $R_{\mathrm{c}}$ must be at least as large as the
largest distance between any two non-zero charges in the system (this
is trivially satisfied if $R_{\mathrm{c}}\geq\sqrt{3}L_{{\rm cell}}$).
In order for the periodic densities not to impinge on each other,
$L_{{\rm pad}}\geq(L_{{\rm mol}}+R_{\mathrm{c}})$ must be satisfied,
where $L_{{\rm mol}}$ is the extent of the system (again, defined
as maximum distance between two non-zero charges) in any Cartesian
direction.}

\end{figure}

The computational overhead of the method during an SCF calculation
compared to the traditional PBC Fourier transform Coulomb approach
consists of three parts: transfer of the calculated density from the
original grid to a larger, padded grid, calculation of the forward
and backwards Fast Fourier Transforms required for the Hartree potential
on the larger grid, and extraction of the calculated potential from
the larger grid back to the original one. The first and last of these
are in general comparatively trivial and take very little time. Performing
the FFT on the larger grid often incurs a considerable slowdown relative
to performing it on the original grid, but nevertheless, generally
speaking, this part of the calculation takes a almost negligible fraction
of the total computational time for large enough systems.

When simulating an isolated object such as a nanocrystal or nanotube
with a high aspect ratio, the geometry of the system requires that
we use a simulation cell that is very long in one dimension (the $x$
direction here) and comparatively small in the other two ($y$ and
$z$). Performing cutoff Coulomb calculations with a spherical cutoff
would rapidly become impractical as the length of the system is increased,
since for a sphere geometry, we would be required to embed the original
cell in a padded cell with all the side lengths $L_{x}\,,L_{y},\, L_{z}>R_{c}$.
In such cases, we need to define a geometry for the cutoff Coulomb
interaction such that the cutoff range can be very long in one direction
and shorter in the other two. One obvious choice for a long, thin
system is to cut off the Coulomb interaction on the surface of a cylinder.
In this case, the integrals required to evaluate the coefficients
are not analytically solvable but can be put in a form amenable to
numerical evaluation. Appendix A gives details on the evaluation of
the Fourier coefficients of the interaction for a cylindrical cutoff.
With an efficient system for evaluating the terms $V_{\mathrm{CC}}(\mathbf{G})$
numerically, the interaction can be calculated rapidly in advance
and reused, and simulations of isolated high aspect ratio systems
can proceed within cells of feasible size.

\section{Minimum Image Convention\label{sec:Minimum-Image-Convention}}

An alternative technique for avoiding periodic interactions is the
class of approaches which includes those of Martyna-Tuckerman \cite{martyna_reciprocal_1999}
and Genovese \emph{et al} \cite{genovese_efficient_2006,genovese_efficient_2007}.
The essence of these, which we will call Minimum Image Convention
(MIC) approaches is that the form of the Coulomb operator is modified
in a way that is still periodic (as this is unavoidable if standard
FFTs are to be used) but which nevertheless removes contributions
from neighbouring cells.

To see how this is achieved, we consider first the Fourier transform
of a function $f(\mathbf{r})$, defined as \begin{equation}
\tilde{f}(\mathbf{G})=\int_{\mbox{all space}}e^{-i\mathbf{G}\cdot\mathbf{r}}\, f(\mathbf{r})\, d\mathbf{r}.\end{equation}
In PBCs, a discrete set of wave vectors $\mathbf{G}$ are used to
expand functions in Fourier space. These wavevectors are chosen by
the requirement that they need to be commensurate with the simulation
cell. Therefore, given the expression for $\tilde{f}(\mathbf{G})$,
the real space representation of the function $f(\mathbf{r})$ under
PBCs is the following: \begin{equation}
f_{{\rm per}}(\mathbf{r})=\sum_{\mathbf{G}\,\,\in\,\,\mbox{cell}}\tilde{f}(\mathbf{G})e^{i\mathbf{G}\cdot\mathbf{r}}\,\,\ .\label{eq:f-per}\end{equation}
This is an exact result and shows that the Fourier representation
of $f(\mathbf{r})$ in the simulation cell is a periodic function
$f_{{\rm per}}(\mathbf{r})$ with the periodicity of the simulation
cell. It is important to notice is that this function is constructed
as a superposition of periodically repeated functions $f(\mathbf{r})$,
one in each cell. This is demonstrated for the example of a Gaussian
function in the top panel of Figure~\ref{fig:ft-mt}, where its resulting
periodic form in one simulation cell is provided, as it would be generated
in real space as a Fourier expansion by Eq.~(\ref{eq:f-per}). This
result implies that periodic interactions are unavoidable if the potential
is constructed by approaches based on Fourier transforms in the standard
simulation cell, as PBCs are implicit in such procedures. However,
MIC approaches are designed to avoid the part of the Coulomb interaction
which produces this undesired long-ranged interaction.

We have implemented the Martyna-Tuckerman approach \cite{martyna_reciprocal_1999},
in which the Fourier method is used to construct not the periodic
function $f_{{\rm per}}(\mathbf{r})$ but the periodic function $f_{{\rm MIC}}(\mathbf{r})$
which results by making $f(\mathbf{r})$ periodic over a single simulation
cell using the MIC \cite{allen_computer_1989}. A similar approach
can also be employed in Quantum Monte Carlo calculations, via the
{}``Model Periodic Coulomb'' approach \cite{williamson_elimination_1997,kent_finite-size_1999}.
The distinction between $f_{{\rm per}}(\mathbf{r})$ and $f_{{\rm MIC}}(\mathbf{r})$
is clarified in Figure~\ref{fig:ft-mt} where the bottom panel demonstrates
the construction of $f_{{\rm MIC}}(\mathbf{r})$ for the example of
a Gaussian function.

To work with this formalism we need to determine the Fourier transform
$\overline{f}(\mathbf{G})$ that will produce the desired $f_{{\rm MIC}}(\mathbf{r})$
\begin{equation}
f_{{\rm MIC}}(\mathbf{r})=\sum_{\mathbf{G}\,\,\,\mbox{cell}}\overline{f}(\mathbf{G})e^{i\mathbf{G}\cdot\mathbf{r}}\,\,\,.\end{equation}
 As this method is intended for dealing with the Coulomb potential,
from now on we will fix the function $f(\mathbf{r})$ to be equal
to $\phi(\mathbf{r})=1/r$ so that we can focus on particular issues
that arise in this case. In determining the form of $\overline{\phi}(\mathbf{G})$
we need to deal with the extra complication of the singularity of
the potential at $\mathbf{r}\rightarrow0$ (short range) and at $\mathbf{G}\rightarrow0$
(long range). The Coulomb potential is partitioned as \begin{equation}
\frac{1}{r}=\frac{{\rm erf}(\alpha r)}{r}+\frac{{\rm erfc}(\alpha r)}{r}=\phi_{{\rm long}}(\mathbf{r})+\phi_{{\rm short}}(\mathbf{r})\,\,\,\,,\label{eq:partition}\end{equation}
 where $\alpha$ is a convergence parameter which determines the region
where the transition from short to long-range terms takes place. Assuming
that the simulation cell is large enough so that $\overline{\phi}_{{\rm short}}(\mathbf{G})\simeq\tilde{\phi}_{{\rm short}}(\mathbf{G})$
only the long range form $\overline{\phi}_{{\rm long}}(\mathbf{G})$
needs to be determined. The desired Fourier transform is expressed
as \begin{eqnarray}
\overline{\phi}(\mathbf{G}) & = & \overline{\phi}_{{\rm long}}(\mathbf{G})+\tilde{\phi}_{{\rm short}}(\mathbf{G})\label{eq:short+long}\\
 & = & [\overline{\phi}_{{\rm long}}(\mathbf{G})-\tilde{\phi}_{{\rm long}}(\mathbf{G})]+[\tilde{\phi}_{{\rm long}}(\mathbf{G})+\tilde{\phi}_{{\rm short}}(\mathbf{G})]\nonumber \\
 & = & \hat{\phi}_{{\rm screen}}(\mathbf{G})+\tilde{\phi}(\mathbf{G})\,\,\,,\label{eq:screen}\end{eqnarray}
 where the explicit expression for $\tilde{\phi}_{{\rm short}}(\mathbf{G})$
is \begin{equation}
\tilde{\phi}_{{\rm short}}(\mathbf{G})=\frac{4\pi}{G^{2}}\left[1-\exp\left(-\frac{G^{2}}{4\alpha^{2}}\right)\right]\;.\label{eq:short-tilder}\end{equation}

Eq.~(\ref{eq:short+long}) can also be further expanded to the form
shown in Eq.~(\ref{eq:screen}) which demonstrates that the MT formalism
is equivalent to augmenting $\tilde{\phi}(\mathbf{G})$ with a {}``screening
potential'' $\hat{\phi}_{{\rm screen}}(\mathbf{G})$ which cuts off
the interactions from the periodic images of the simulation cell.
In practice, we compute $\overline{\phi}(\mathbf{G})$ according to
Eq.~(\ref{eq:short+long}) and we distinguish two cases: $\mathbf{G\neq0}$
and $\mathbf{G=0}$, which must be treated separately.

The function $\overline{\phi}_{{\rm long}}(\mathbf{G})$ for $\mathbf{G\neq0}$
is obtained as \begin{equation}
\overline{\phi}_{{\rm long}}(\mathbf{G})=\int_{{\cell}}e^{-i\mathbf{G}\cdot\mathbf{r}}\frac{{\rm erf}(\alpha r)}{r}d\mathbf{r}\;,\label{eq:long-over}\end{equation}
 where the above integral is computed as a sum over the simulation
cell grid points as this is an exact expression for the wavevectors
$\mathbf{G}$ which are commensurate with the simulation cell. The
above expression is the desired one as the term ${\rm erf}(\alpha r)/r$
does not contain contributions from periodic images. It also does
not contain a singularity at $r=0$ so the evaluation of this integral
poses no difficulties. The complete expression for $\overline{\phi}(\mathbf{G})$
is obtained as the sum of the terms Eq.~(\ref{eq:short-tilder})
and Eq.~(\ref{eq:long-over}). 

To find the $\mathbf{G=0}$ term, we need to consider the limit of
Eq.~(\ref{eq:short-tilder}) as $\mathbf{G}$ goes to zero \begin{eqnarray}
 &  & \lim_{\mathbf{G}\rightarrow0}\tilde{\phi}_{{\rm short}}(\mathbf{G})=\nonumber \\
 & = & \lim_{\mathbf{G}\rightarrow0}\frac{4\pi}{G^{2}}\left[1-\left(1-\frac{G^{2}}{4\alpha^{2}}+\frac{G^{4}}{8\alpha^{4}}+\cdots\right)\right]\nonumber \\
 & = & \frac{\pi}{\alpha^{2}}\label{eq:short-zero}\end{eqnarray}
and taking this into account, Eq.~(\ref{eq:short+long}) becomes
\begin{eqnarray}
\overline{\phi}(\mathbf{0}) & = & \overline{\phi}_{{\rm long}}(\mathbf{0})+\tilde{\phi}_{{\rm short}}(\mathbf{0})\nonumber \\
 & = & \int_{{\cell}}\frac{{\rm erf}(\alpha r)}{r}d\mathbf{r}+\frac{\pi}{\alpha^{2}}\;,\end{eqnarray}
where the integral in the above expression is again evaluated as a
sum over the simulation cell grid points as the integrand does not
contain a singularity at $r=0$. 

In order to use the MT potential in practical calculations, we need
to ensure that appropriate conditions are obeyed as regards the relative
sizes of the simulated molecule and the simulation cell. From the
example in the bottom panel of Figure~\ref{fig:ft-mt} we can see
that the length that a simulation cell can have in any direction needs
to be at least twice the length of the molecule being simulated. In
the opposite case unphysical interactions will be introduced as some
charges on the molecule will be experiencing the Coulomb potential
from other parts of the molecule (as they should) while other charges
will experience the potential from a periodic image (which they should
not).

In our implementation the Hartree potential is generated in reciprocal
space from the electronic density as a product with the Fourier transform
of MT potential $\overline{\phi}(\mathbf{G})$ \begin{equation}
\overline{V}_{\mathrm{H}}(\mathbf{G})=\overline{\phi}(\mathbf{G})\tilde{n}(\mathbf{G})\;.\end{equation}
 In a similar way, the local pseudopotential is obtained in reciprocal
space as a sum of short and long range terms \begin{equation}
\overline{V}_{{\rm locps}}(\mathbf{G})=\overline{V}_{{\rm locps,short}}(\mathbf{G})+\overline{V}_{{\rm locps,long}}(\mathbf{G})\,\,\,\,.\end{equation}
 For an ion with charge $-Z$, (following the established electronic
structure theory convention of taking the ionic potential as negative),
the periodic Coulomb component is subtracted from the pseudopotential
to obtain its short range part \begin{equation}
\overline{V}_{{\rm locps,short}}(\mathbf{G})=\tilde{V}_{{\rm locps}}(\mathbf{G})+Z\tilde{\phi}(\mathbf{G})\end{equation}
 and the long range part is obtained as the MIC Coulomb interaction
\begin{equation}
\overline{V}_{{\rm locps,long}}(\mathbf{G})=-Z\overline{\phi}_{{\rm long}}(\mathbf{G})\;.\end{equation}
 Finally the core-core interaction energy is obtained as a Coulombic
sum between point charge interactions in the simulation cell according
to Eq.~(\ref{eq:eionion}).

Genovese \textit{et al.} \cite{genovese_efficient_2006,genovese_efficient_2007}
proposed an approach that is rather similar in principle but in practice
has some different properties. They described a wavelet-based approach
to calculating the MIC Coulomb interaction. The charge density is
expanded using interpolating scaling functions \cite{deslauriers_symmetric_1989}
of order $m$ (typically $m=14$). This guarantees that when a known
continuous charge density is represented, the first $m$ moments are
preserved. Although most practical methods do not attempt to represent
given continuous charge densities, this approach is useful when using
pseudopotentials of the form proposed by Goedecker \textit{et al.}~\cite{goedecker_separable_1996}.
The representation of the Coulomb operator is made separable by employing
an expansion in terms of Gaussians \cite{saito_multiresolution_1993}.
The resulting one-dimensional integrals can be calculated to machine
precision by exploiting the refinement relation of scaling functions
and then tabulated for future use. The necessary convolution to obtain
the Hartree potential requires FFTs on a grid that is doubled in each
dimension to avoid spurious periodic interactions, but this can be
performed without additional computational effort by modifying the
FFT algorithm to exploit the fact that the charge density is zero
on the additional grid points. This latter optimisation would also
benefit the Cutoff Coulomb approach. A representation of the Hartree
potential arising from the MIC Coulomb potential results that is essentially
exact.

\section{Open Boundary Conditions\label{sec:Open-Boundaries}}

The final possibility we will consider is to change not the form of
the interactions, but that of the boundary conditions. A careful recasting
of the electrostatic terms in the Kohn-Sham energy functional allows
us to use a form suitable for calculation with Open Boundary Conditions
(OBCs). This is achieved by replacing the reciprocal-space evaluation
of the core-core, Hartree and local pseudopotential energy terms by
calculations performed in real space, which assume no periodicity
of the system.

The core-core interaction energy is calculated as a Coulombic sum
of the interaction energies of point charges as in Eq.~(\ref{eq:eionion}).
We describe in Appendix B how the local pseudopotential $\vlocr$
can be calculated in real space.

\begin{figure}[h]
 \includegraphics[width=0.65\columnwidth]{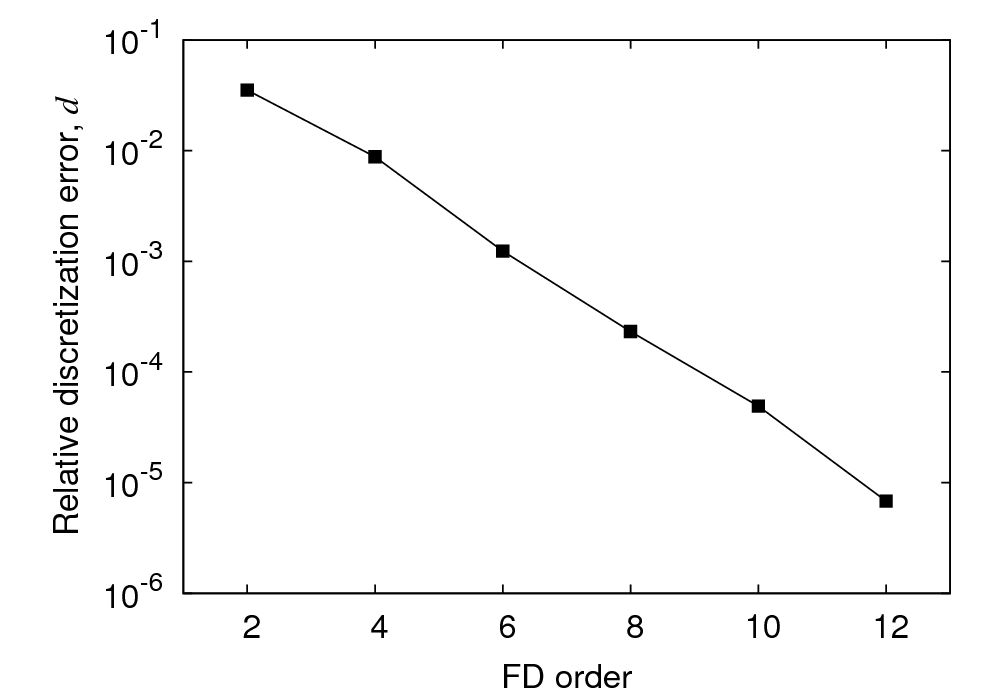} \caption{\label{fig:PBC_mg_discrepancy} Relative discretization error Eq.~(\ref{eq:d})
in the Hartree energy \vs{} the order of the finite differences
used in the defect correction of the second-order solution, on the
example of aspartate. An order of 2 corresponds to the uncorrected
solution. Smeared ions were used.}

\end{figure}

The Hartree potential $V_{\mathrm{H}}(\mathbf{r})$ is obtained by
solving the Poisson Eq.~(\ref{eq:Poisson_2}) in real space. The
multigrid method \cite{Brandt-1977-333} represents an efficient approach
for solving for the potential, given the charge density sampled on
a regular grid and Dirichlet boundary conditions on the faces on the
simulation cell, $\partial\cell$. By using a hierarchy of successively
coarser grids along with interpolation and restriction operators to
transfer the problem between the grids, the multigrid approach addresses
the problem of critical slowing down that plagues stationary iterative
methods \cite{Hatems_thesis}. For a more thorough discussion of the
approach the reader is referred to Refs.~\cite{beck_real-space_2000,Trottenberg-2001-631,Brandt-1977-333}.
In the simplest approach, second-order finite differences (FDs) are
used to approximate the Laplacian in Eq.~(\ref{eq:Poisson_2}). However,
there is evidence \cite{PhysRevLett.72.1240,Hatems_thesis} that this
is not sufficiently accurate for DFT calculations. One way to assess
the accuracy of the solution is by comparing the values of two expressions
for the Hartree energy, namely \begin{equation}
E_{\textrm{H}}^{0}=\frac{1}{2}\int_{\cell}V_{\mathrm{H}}(\mathbf{r})n(\mathbf{r})d\mathbf{r}\label{eq:EH0}\end{equation}
 and \begin{equation}
E_{\textrm{H}}^{1}=\frac{1}{8\pi}\left[\int_{\cell}{\left(\nabla{}V_{\mathrm{H}}(\mathbf{r})\right)}^{2}d\mathbf{r}-\oiint_{\partial\cell}V_{\mathrm{H}}(\mathbf{r})\nabla{}V_{\mathrm{H}}(\mathbf{r})d\mathbf{S}\right].\label{eq:EH1}\end{equation}
 The relative discretization error, defined as \begin{equation}
d=\left|\frac{E_{\textrm{H}}^{1}-E_{\textrm{H}}^{0}}{E_{\textrm{H}}^{0}}\right|\label{eq:d}\end{equation}
 can then serve as a measure of the inaccuracy of the solution. Figure~\ref{fig:PBC_mg_discrepancy}
shows how this error is unacceptably large when a second-order solver
is used. The problem can be addressed by employing high-order defect
correction, where higher-order finite differences are used to iteratively
correct the solution obtained with a second-order solver \cite{Schaffer-1984-89}.
In this way the discretization error can be systematically reduced
(Figure~\ref{fig:PBC_mg_discrepancy}) with moderate computational
cost. No changes to the second-order solver are necessary. The computational
cost of the multigrid approach scales linearly with the volume of
the simulation cell, albeit with a large prefactor.

The multigrid method does not rely on any particular form of the Dirichlet
boundary conditions specified on $\partial\cell$, however, to obtain
a potential consistent with the OBCs used for the remaining energy
terms, these should be \begin{equation}
V_{\mathrm{H}}(\mathbf{r})=\int_{\cell}\frac{n(\mathbf{r}')}{\left|\mathbf{r}-\mathbf{r}'\right|}d\mathbf{r}'\textrm{ for }\mathbf{r}\in\partial\cell\;.\label{eq:OBC}\end{equation}
 Although the evaluation of the boundary conditions is straightforward,
it is computationally costly, scaling unfavourably as $O(L^{2}V)$,
which, for localised charge, implies $O(L^{2}N)$. To ameliorate this
problem, a suitable coarse-grained approximation can be used instead
of $n(\bvec{r'})$. Combined with evaluating Eq.~(\ref{eq:OBC})
only for a subset of points in $\partial\cell$ and using interpolation
in between, this leads to a reduction of the computational effort
by 3-4 orders of magnitude, which brings it into the realm of feasibility.

The smeared-ion formalism \cite{Scherlis-2006-074103}, where the
total energy is rewritten by adding and subsequently subtracting Gaussian
charge distributions centred on the cores, can be used in conjunction
with the multigrid approach. In this case, the Poisson equation, Eq.~(\ref{eq:Poisson_2}),
is solved for the electrostatic potential generated by the \emph{total}
charge density (due to electrons and smeared ions). As the cores neutralize
a significant fraction of the electronic charge, the magnitude of
the relevant quantities (charge density, potential) is smaller. Assuming
the relative error incurred by the multigrid remains the same, this
has the advantage of reducing the absolute error. The use of smeared
ions, however, introduces approximations of its own. For a more detailed
discussion of smeared ions the reader is referred to Ref.~\cite{Scherlis-2006-074103}.
We shall evaluate the approach with and without smeared ions.

\section{Convergence Properties\label{sec:Convergence}}

\subsection{Small Molecular Systems}

\begin{figure*}
\includegraphics[width=0.8\columnwidth]{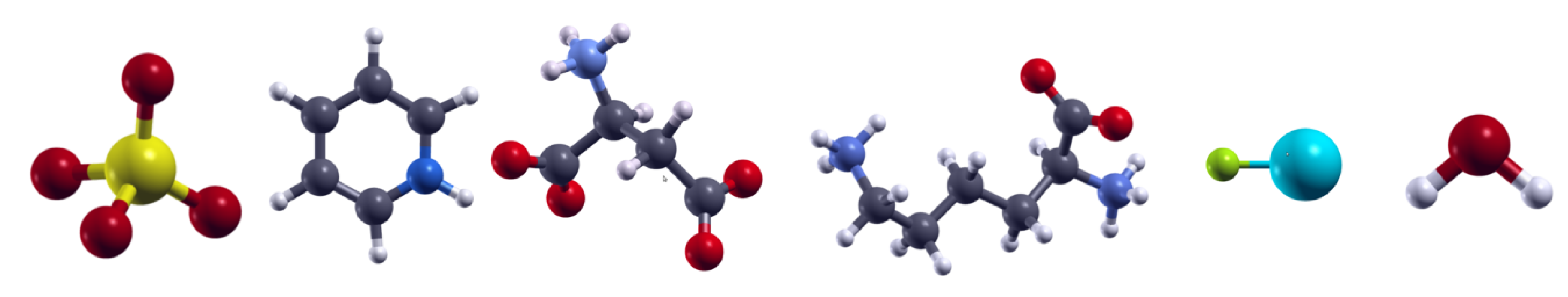}

\caption{\label{fig:small_molecules}(Color online) Small molecules for initial
tests, covering anions and cations and species with dipole moments.
From left to right: phosphate, pyridinium, aspartate, lysine, potassium
chloride and water.}

\end{figure*}

We test these methods first on small-scale, simple systems to demonstrate
their equivalence in the limit that all relevant parameters are accurately
converged. For this, we select a test set of small ions molecules:
a phosphate ion (PO$_{4}^{\phantom{4}3-}$), pyridinium (C$_{5}$NH$_{6}$)$^{1+}$,
the amino-acid salt aspartate with a charge of $-1e$, and the amino
acid lysine with a charge of $+1e$, the neutral molecules water (H$_{2}$O)
and potassium chloride (KCl). In this set, we have thus included two
cations, two anions and neutral molecules with a relatively low and
a very high dipole moment, respectively. Clearly, these small molecules
are unchallenging calculations for linear-scaling DFT, of a size below
the onset of any linear-scaling behaviour, but they serve to illustrate
the main convergence issues in a controllable way, since it is here
possible to make the simulation cell very much larger than the molecule,
within feasible computational memory requirements. Illustrations are
shown in Figure~\ref{fig:small_molecules} of this test set.

Each molecule was simulated in a cubic simulation cell initially of
size 32.5$a_{0}$, with a grid spacing of 0.5$a_{0}$ (equivalent
to an energy cutoff of around 827eV), and with all NGWF radii set
to 7.0$a_{0}$. The density kernel was not truncated (all elements
allowed to be nonzero) as the systems are too small for meaningful
truncation. Norm-conserving pseudopotentials are employed for all
the ions included here, and exchange-correlation is described by the
PBE functional. We choose, throughout this work, to examine the convergence
of the total energy, because although in practice one is most often
interested in a quantity derived from it, such as formation or binding
energies, the finite size errors made in total energies due to monopole
or higher charges cannot be expected to cancel between (for example)
reactant and product states, so convergence of the total energy must
be obtained individually for each system.

\subsection{Convergence}

\begin{figure*}
\includegraphics[clip,width=0.4\paperwidth]{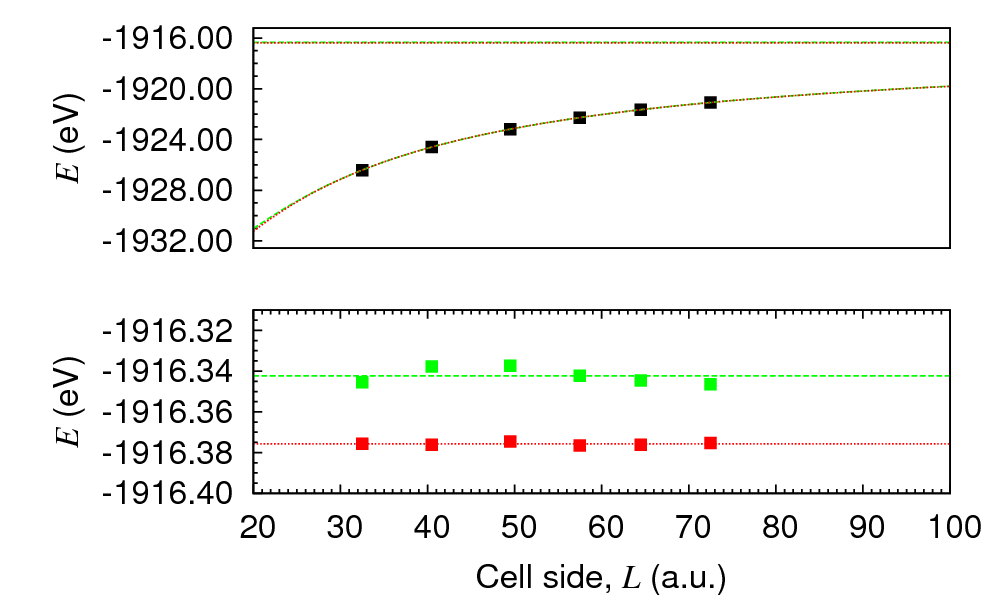}\includegraphics[width=0.4\paperwidth]{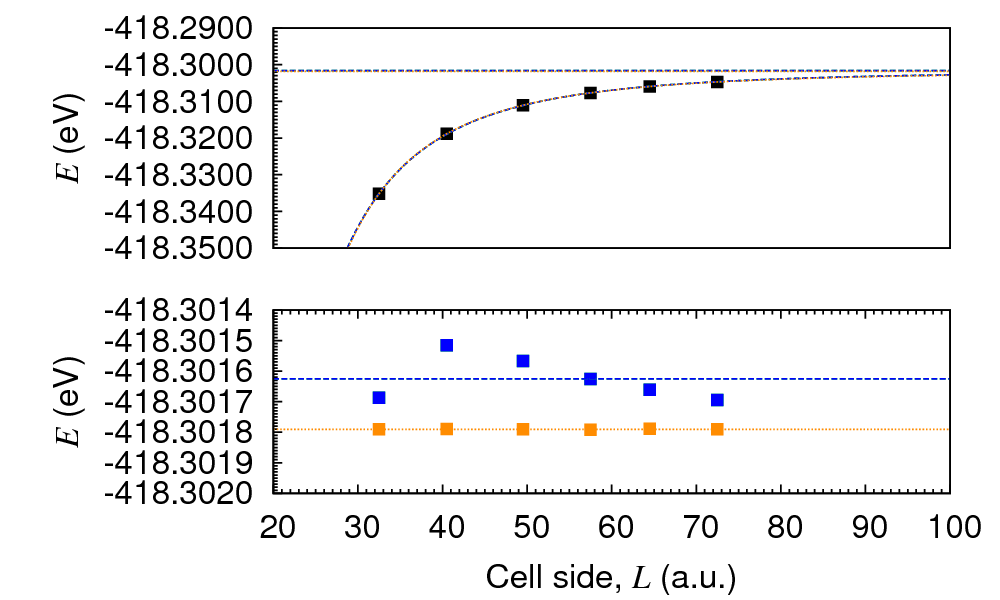}\caption{\label{fig:mp_small_molecules}Convergence of total energy with simulation
cell size for a monopolar system (PO$_{4}^{3-}$, left) and a dipolar
system (KCl, right), showing the uncorrected results (black), and
different forms of Makov-Payne correction: a) red: $E(L)=E_{0}+q^{2}\alpha/2L+B/L^{3}$;
b) green: $E(L)=E_{0}+A/L+B/L^{3}+C/L^{5}$; c) blue: $E(L)=E_{0}+B/L^{3}$;
d) orange: $E(L)=E_{0}+B/L^{3}+C/L^{5}$.}

\end{figure*}

First, we examine the option of extrapolation to infinite size from
calculations performed under PBCs. The black diamonds in Figure~\ref{fig:mp_small_molecules}
show the uncorrected total energy of each of the six molecular species,
calculated under PBCs. According to Makov and Payne \cite{makov_periodic_1995},
the total energy as a function of box size can be expected to behave
approximately as\begin{equation}
E=E_{0}-\frac{q^{2}\alpha}{2L}-\frac{2\pi qQ}{3L^{3}}+O(L^{-5})\;\label{eq:MP_corr}\end{equation}
in a cubic simulation cell of side $L$, where $q$ is the total charge,
$Q$ is the quadrupole moment, and $\alpha$ is the Madelung constant,
where for cubic cells $\alpha\simeq2.837$. They suggest an approximate
correction scheme based on removing the leading order $L$-dependent
term. However, there are two options for going about this in practice.

Direct calculation of quadrupole moment $Q$ from the density is problematic
and a more reliable approach is to set the monopole charge $q$ according
to the known charge and then fit $E_{0}$ and $Q$ to data using a
least-squares fit to data at multiple values of $L$. This produces
the dotted black line in each figure. Alternatively, one could take
into account that for a cell containing a molecule which is to some
extent extended and may be somewhat polarisable, the mean dielectric
constant is not equal to precisely unity. One could therefore also
allow the coefficient of the $1/L$ term to vary freely, and allow
an $O(L^{-5})$ coefficient as well. Examining Figure~\ref{fig:mp_small_molecules}
we see that the finite size error for those species with a monopole
charge follows $E(L)=E_{0}+O(L^{-1})$ fairly well as expected. The
species with only a dipole moment (not a monopole) display a much
weaker effect, which behaves as $E(L)=E_{0}+O(L^{-3})$. However,
as the charge distribution varies with $L$, and so the coefficient
$Q$ in Eq.~(\ref{eq:MP_corr}) depends weakly on $L$, the fit to
the Makov-Payne form is not exact. Nevertheless, in the small \emph{charged}
molecules used here, the fitted Makov-Payne correction achieves a
fairly well-defined correction to the total energy, aligning each
individual energy to the extrapolated infinite-cell-size limit even
for smaller cells, producing a good fit. However, the extra freedom
allowed by varying $q$ or introducing $O(L^{-5})$ terms are seen
to produce a less useful extrapolation, by fitting to noise. This
can only be seen for sure by comparing to the known answer obtained
under one of the correction schemes as seen below.

The effect of self-consistency in these small systems is not very
strong: that is to say, the rearrangement of the charge due to the
influence of the potential from neighbouring images of the cell is
not very great. Henceforth, for Makov-Payne results, we will show
the \emph{corrected} result $E_{\mathrm{PBC}}(L)-E_{\mathrm{MP}}(L)+E_{0}$,
where $E_{\mathrm{MP}}(L)$ is the appropriate Makov-Payne choice,
as this result falls on a comparable scale to the results for the
other schemes, enabling visual comparison.

Within the cutoff Coulomb approach, we can individually vary the size
of the original cell, the size of the padded cell, and the cutoff
radius of the interaction. We note that the results obtained are converged
to less than 1$\mu$eV/atom once the extent of the density of the
molecule is less than that of the original cell. Since the interaction
is constructed in reciprocal space but has a sharp cutoff in real
space at $R_{c}$, care must be taken to include sufficient padding
that the cutoff falls within a vacuum region, and the region of `ringing'
induced by the cutoff is at least 5-10$a_{0}$ from any significant
values of nonzero density. Once this is achieved, residual variation
of the result with $R_{c}$ is well below 1$\mu$eV/atom.

For the MIC approach, we obtain near-identical results for our implementation
of the Martyna-Tuckerman approach as compared to our implementation
of the approach of Genovese \emph{et.~al.} We thus only show the
MT approach henceforth, which was rather easier to parallelise in
the current methodology, even though the Genovese\emph{ et.~al.}
approach is technically more sophisticated and requires less computational
effort overall due to the lower padding requirements. Martyna and
Tuckerman note that to obtain accurate reciprocal-space representation
of the MIC Coulomb potential, a smaller grid spacing is sometimes
required compared to the requirements of a comparable PBC calculation.
Alternatively, one can represent just the density and potential on
a finer grid. Taking the latter approach, we compared grid spacings
$2.0\times$, $2.5\times$ and $3.0\times$ the underlying psinc grid
for representation of the density and potential. While the results
do show minor variations (from 20 to 100 $\mu$eV/atom depending on
the system), this variation is present to the same extent also in
PBC calculations so should not be attributed to the MT approach itself
--- rather it is thought to result from changing the grid in discrete
evaluation of the XC energy integral. We thus employ the standard
$2.0\times$fine grid spacing throughout the rest of this work.

\begin{figure*}
\includegraphics[clip,width=0.35\paperwidth]{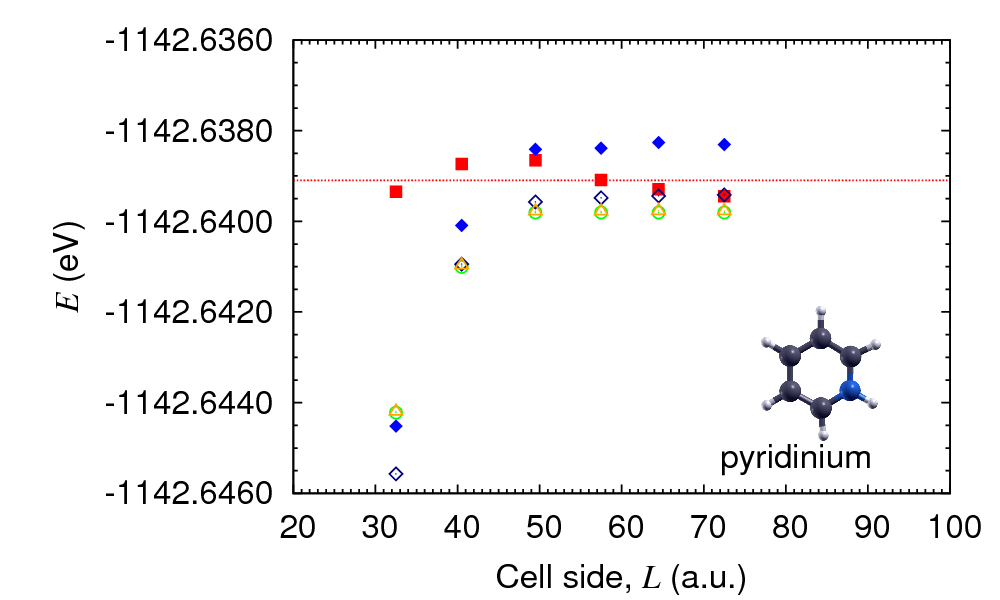}\includegraphics[width=0.35\paperwidth]{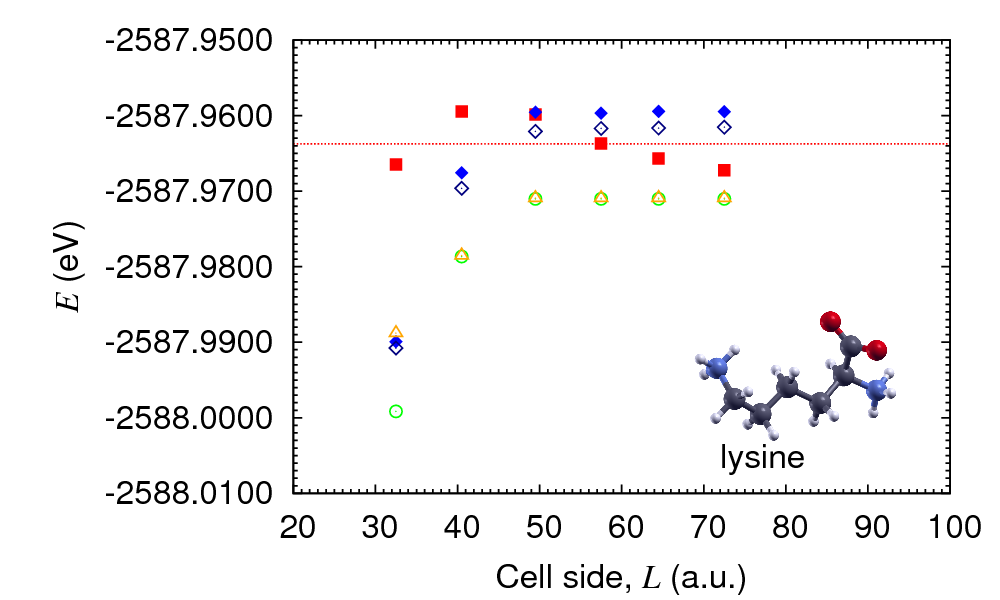}

\includegraphics[width=0.35\paperwidth]{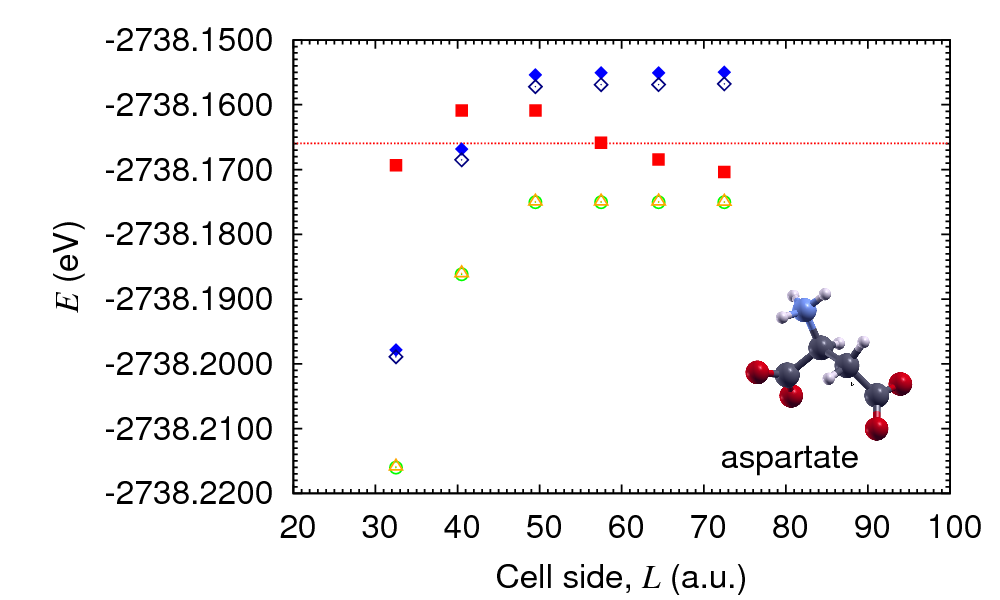}\includegraphics[width=0.35\paperwidth]{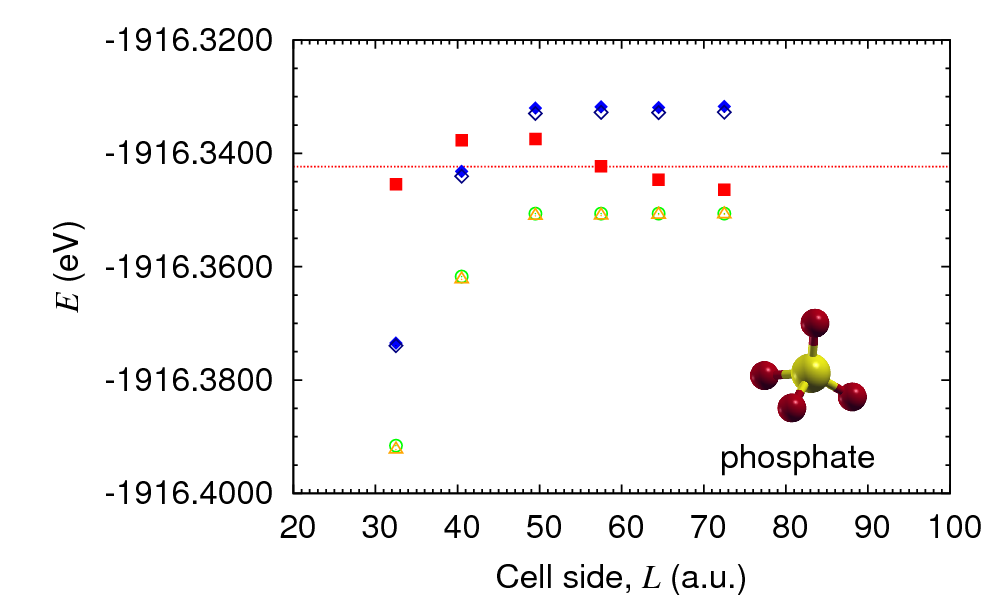} 

\includegraphics[width=0.35\paperwidth]{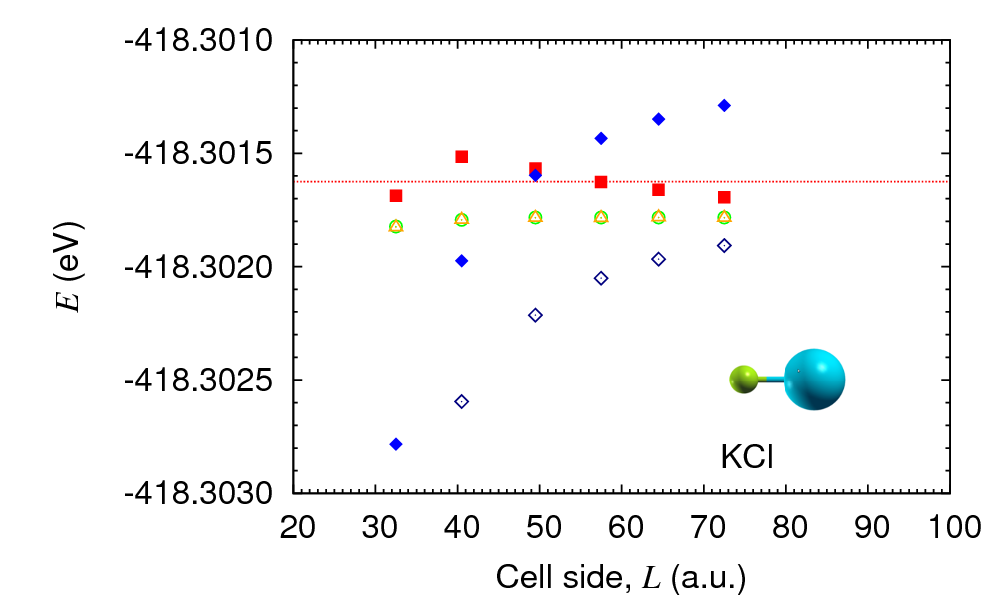}\includegraphics[width=0.35\paperwidth]{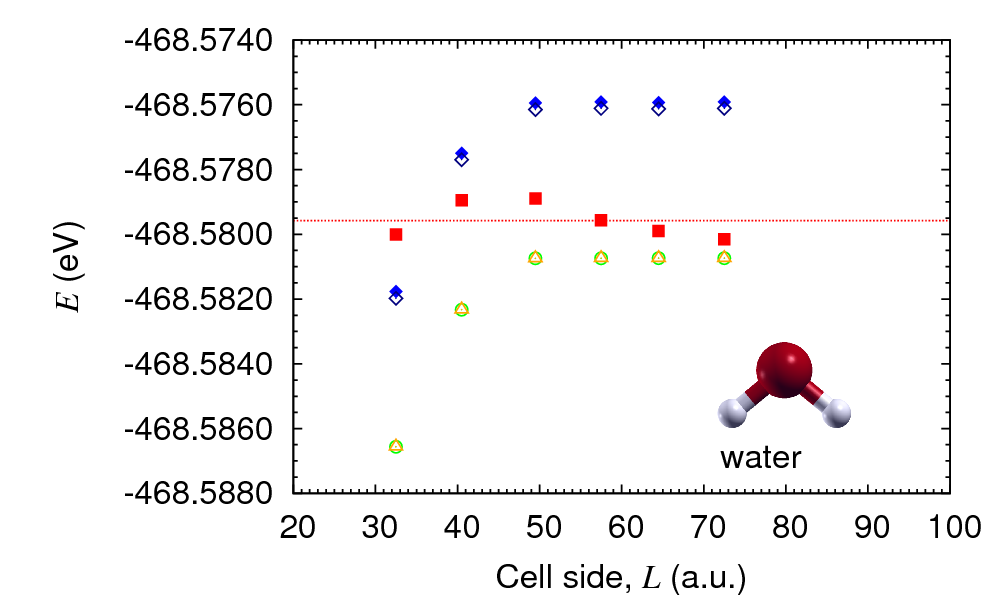}\caption{\label{fig:te_small_molecules}Convergence of total energy with simulation
cell size for test molecules, using: Cutoff Coulomb (green circles),
Martyna-Tuckerman (orange triangles), OBCs (blue diamonds, filled
when smeared ions were used), and MP-corrected PBCs (red squares).
CC and MT results rapidly approach the same converged answer once
the size of the cell is greater than the extent of the density. This
converged result agrees well with the trend of the MP-corrected lines.
The OBC results are offset by a constant due to the approximations
involved in the smeared-ion representation and by an error whose magnitude
increases with the box size (particularly evident for KCl) as a consequence
of the numerical evaluation of the real-space pseudopotential (see
Appendix B).}

\end{figure*}

We show in Figure~\ref{fig:te_small_molecules} the total energy
of the test systems evaluated all the above methods. The CC results
use the spherical cutoff of Eq.~(\ref{eq:v_CC_sphere}). The results
for the CC and MIC methods converge rapidly with system size to effectively
the same value. In very small simulation cells, below 42$a_{0}$,
the extent of the `FFT box' --- and thus the total extent of the charge
distribution --- is the same as that of the simulation cell. In such
cases, the simulation cell contains very small contributions to the
total electron density that wrap through the periodic boundaries.
Therefore, even the correction schemes do not fully account for the
absense of periodic interactions, and a quite strong dependence on
$L$ at very small $L$ is seen. However, as soon as the simulation
cell is large enough that the density is contained fully within one
cell, the result is beyond that point entirely converged with system
size and independent of $L$.

However, the OBC calculations evidently produce results of a somewhat
lower accuracy. For these results, several distinct sources of inaccuracy
can be distinguished. First and foremost, the calculation of the local
pseudopotential under OBCs is performed numerically and the associated
error increases with the size of the simulation cell. The reasons
behind this are explained in detail in Appendix B. For the systems
and box sizes encountered here, the magnitude of this error is $20-200\mu$eV/atom,
thus it is only apparent in the plots for KCl, where the magnification
is the highest. Second, the use of a multigrid approach to solve Eq.~(\ref{eq:Poisson_2})
introduces a discretization error. The magnitude of this error, however,
can be easily made negligible by employing high-order defect correction,
and introducing smeared ions, as explained earlier in section~\ref{sec:Open-Boundaries},
cf.~Fig.~\ref{fig:PBC_mg_discrepancy}. Third, there are approximations
involved in the generation of boundary conditions Eq.~(\ref{eq:OBC})
for the solution of Eq~(\ref{eq:Poisson_2}). In our implementation
we coarse-grain charge densities (electronic when smeared ions are
not used, or total when using smeared ions) represented on a grid
by replacing cubic blocks of $p\times p\times p$ gridpoints with
a single point charge located at the centre of charge of the block
(thus, in general, not at a gridpoint). This is only done when evaluating
the integral in Eq~(\ref{eq:OBC}). With $p=5$ (used throughout
this work) the the prefactor for the calculation of the boundary conditions
is reduced $125$-fold, whereas the associated error in the energy
was less than $75\mu$eV/atom in the worst case (PO$_{4}^{\phantom{4}3-}$
in the smallest box) and diminished quickly with increasing box sizes.
For neutral systems, even with high dipole moments, this error was
less than $6\mu$eV/atom, again quickly diminishing with the box size.

Finally, the introduction of smeared ions \cite{Scherlis-2006-074103}
also affects the obtained energies, as evidenced by the near-constant
shifts between the results with and without smeared ions, observed
in the plots. The error incurred by using smeared ions is due to the
fact that certain terms in the formalism (\eg~the self-interaction
of every smeared ion) are calculated analytically, whereas other terms
(\eg~the local pseudopotential energy) are calculated by integrating
the relevant quantities on a grid. Thus, the terms that are meant
to cancel only do so in the limit of an infinitely fine grid. For
the systems discussed here, the residual error is $100-300\mu$eV/atom,
outweighing the reduction in the other sources of error that smeared
ions bring about -- it is apparent from the plots that the calculations
would be more accurate without smeared ions. Smeared ions find use
in the context of implicit solvent calculations \cite{dziedzic_minimal_2011},
as they allow the dielectric continuum to polarise in response not
only to the electronic density, but also to the density of the smeared
cores. For calculations in vacuum involving the systems of interest
here, their introduction negatively impacted accuracy.

Overall, one can conclude from these tests that in small systems,
both the CC and MIC methods can be used with confidence, once the
system size is large enough that the charge density is fully contained
within the appropriate box. Extrapolation-based techniques can correct
energies to comparable accuracy, but should be used with care and
the use of excessive variational freedom in the parameters tends to
worsen results. Finally, when using OBCs, the energy is actually expected
to very slowly diverge with the size of the simulation cell, due to
the inaccuracies involved in the evaluation of the local pseudopotential.
This effect, compounded by the near-constant shift due to the use
of smeared ions means that OBC results should only be compared against
other OBC results rather than results from PBC calculations.

\section{Large Systems and Computational Overhead\label{sec:Overhead}}

To validate and compare these methods in a more realistic setting,
it is necessary to examine their performance in larger-scale systems
more typical of the real applications of linear-scaling DFT. These
will often behave very differently from very small systems, because
it is usually impossible to perform the calculations in a simulation
cell where the dimensions of the cell greatly exceed that of the molecule
or nanostructure. Furthermore, the scaling of the computational effort
with system size may be very different as the balance of time spent
in different parts of the calculation changes with the number of atoms.

\begin{figure}
\includegraphics[width=0.38\columnwidth]{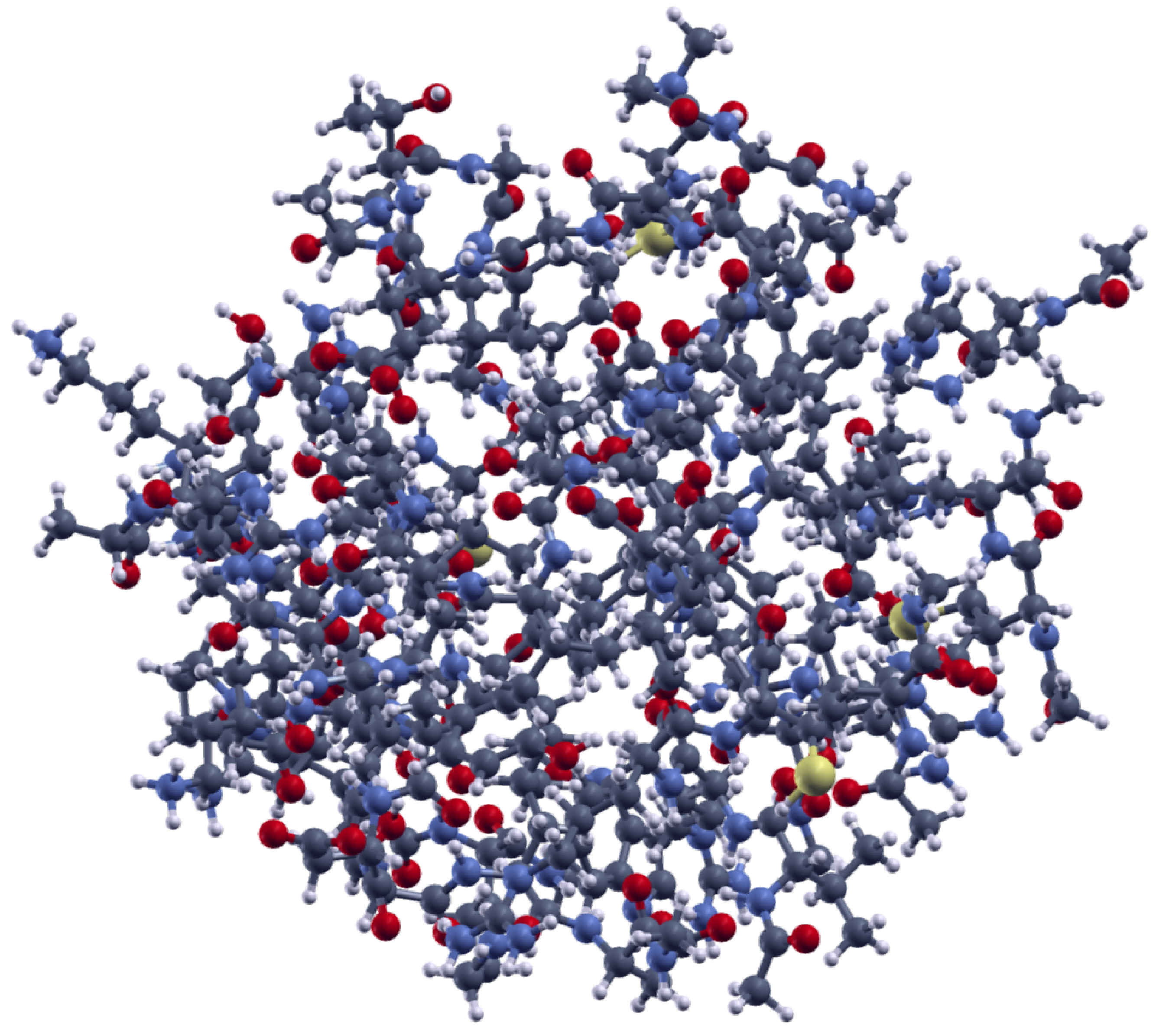}

\includegraphics[width=0.7\columnwidth]{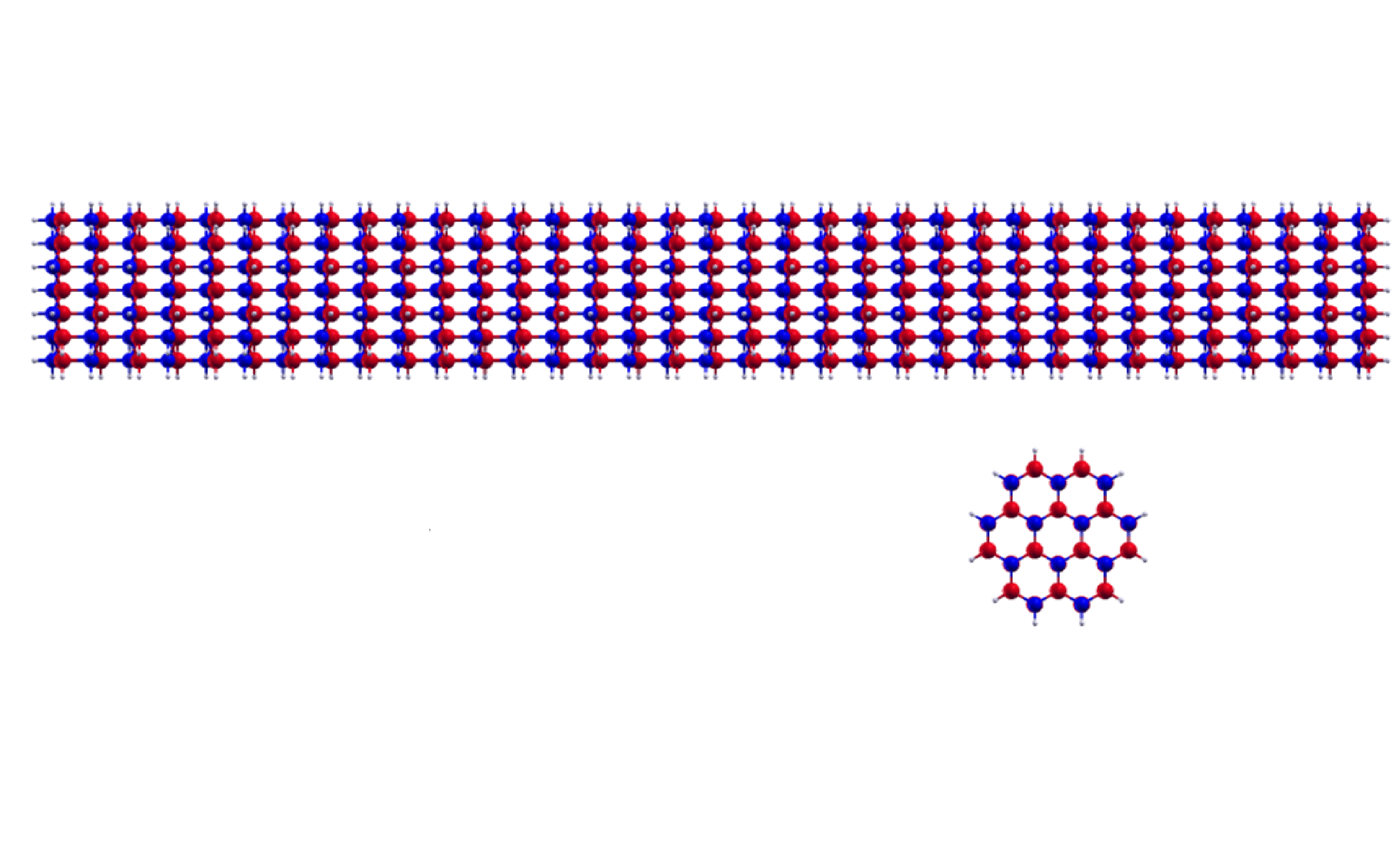}

\caption{\label{fig:large_systems}Illustration of test set of large systems
(a) 1234-atom fragment of the L99A/M102Q mutant of the T4 Lysozyme
protein (b) Wurtzite-structure GaAs Nanorod of 1284 atoms, with hydrogen
atoms terminating dangling bonds.}

\end{figure}

To demonstrate the accuracy of the methods, and the computational
overhead and the scaling of each of these approaches, we have simulated
two fairly large systems, each comprising around 1250 atoms, which
for these systems is well above the threshold at which linear-scaling
methods offer a clear advantage in terms of reduced computational
effort over comparable traditional DFT approaches. These systems are:
a fragment of the L99A/M102Q mutant of the T4 Lysozyme protein \cite{boyce2009,dziedzic_minimal_2011},
and a nanocrystal of gallium arsenide in the wurtzite structure, with
hydrogen termination \cite{avraam_factors_2011}. Figure~\ref{fig:large_systems}
illustrates these systems.

The protein fragment has a high net charge of $+7e$ as a result of
the protonation state of its residues at normal pH, and hence displays
a strong finite size effect on the total energy if periodic boundary
conditions are employed. This makes it difficult to calculate meaningful
binding energies of ligands to its polar binding site. The distribution
of the net charge is largely determined by the functional groups included
and to a great extent it can be viewed as localised on these groups,
so it is not expected to depend strongly on the system size: to a
reasonable approximation we can treat the density of this system as
fixed when we vary the size of the simulation cell.

The GaAs nanorod, on the other hand, has no net charge, but the underyling
wurtzite structure, with no inversion symmetry, means that when truncated
at each end of the rod along the $c$-axis, Ga and As faces are exposed
on opposite ends of the rod. No matter how the surface is terminated
(in the case studied here, all dangling bonds are saturated with hydrogen),
there will be some form of charge transfer between the ends and a
dipole moment along the $c$-axis will result. If such a rod is simulated
in a box of size comparable to the rod itself under PBCs, then the
rod is effectively surrounded by an inifinite array of replicas, producing
a very different electric field from that of an isolated rod. Indeed,
unless the box is very large along all axes, the Ga-terminated end
of the rod will be in closer proximity to the As-terminated ends of
rods in adjacent cells than to the As-terminated end on the on the
same rod (and vice versa), and the rod is strongly polarisable. This
is clearly a very different situation from the ideal situation many
correction methods assume, of a strongly localised fixed charge distribution
in a box considerably larger than the charge distribution itself.
Because the magnitude of the dipole moment depends sensitively on
the balance of charge distribution and the density of states at the
polar surfaces of the rod, its value can be affected by the field
created by the charge distribution of periodic images of the rod,
bringing self-consistent effects into play.

To perform these large simulations, we use in both cases a grid spacing
of 0.5$a_{0}$, equivalent to a plane-wave cutoff of around 827eV,
and standard well-tested norm-conserving pseudopotentials for each
species. For the protein system, four NGWFs of radius 7.0$a_{0}$
were placed on each C, N, O and S atom, and one on each H atom. For
the nanorod, larger NGWFs were required to achieve good convergence,
so $R_{\phi}=10a_{0}$ was used, with four NGWFs on Ga and As and
one on H.

\begin{figure}
\includegraphics[width=1\columnwidth]{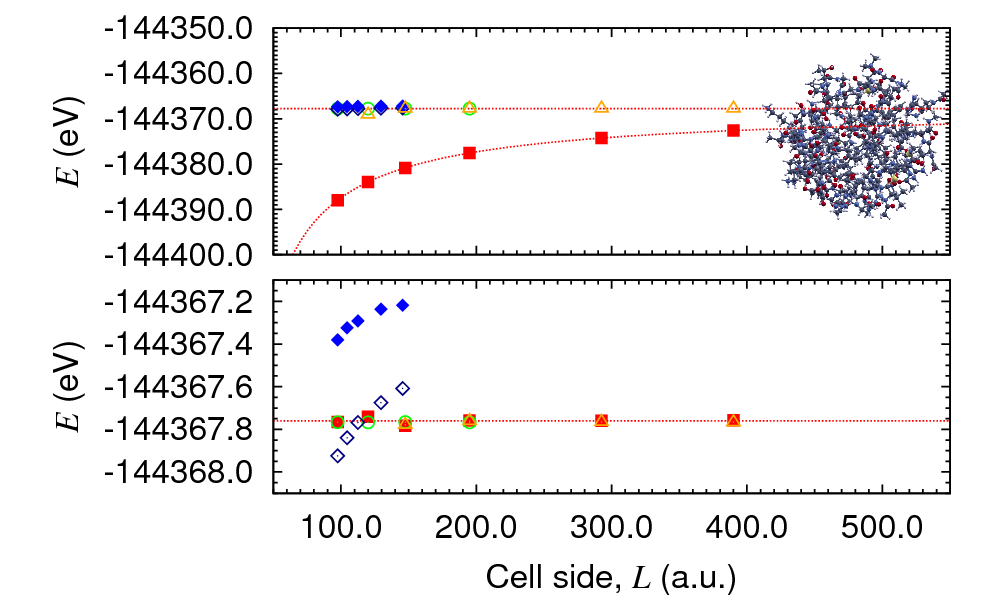}

\caption{\label{fig:catechol-PBC_CC}Convergence of total energy with cell
size for T4 lysozyme fragment, showing results for Cutoff Coulomb
(green circles), Minimum Image Convention (orange triangles), Open
Boundary Conditions (blue diamonds, filled when smeared ions were
used) and Periodic Boundary Conditions (red squares) corrected using
the Makov-Payne form (a) fitted by least-squares fitting to $E_{0}$
and $B$.}

\end{figure}

The total energy of the the protein fragment as a function of supercell
side length is shown in Figure~\ref{fig:catechol-PBC_CC}. We use
a series of cells up to $L=400a_{0}$ in size, so as to be able to
accurately extrapolate to $L\rightarrow\infty$. We see that on the
larger scale (top figure), the results for all three methods lie on
apparently the same line, agreeing with the extrapolation of the Makov-Payne
form to large $L$. However, zooming in reveals two significant details:
firstly, there remains considerable residual variation in the Makov-Payne
corrected results, which do not converge to better than 0.05eV until
$L=200a_{0}$. When they do so, they agree well with the MP extrapolation.
The OBC result suffers from considerable residual error, mostly due
to the approximations involved in the evaluation of the local pseudopotential,
which for the smaller box sizes cancels out, to a degree, with the
error due to the approximations in the construction of the boundary
conditions, but for larger boxes causes the energy to very slowly diverge.
The almost constant shift in energy incurred by the use of smeared
ions is approximately $400\mu$eV/atom. The CC and MT results agree
very well with each other, to around the 1meV level. We conclude that
for monopolar systems with an approximately fixed charge distribution,
the CC and MT methods are both reliable as long as the cell is made
large enough for the conditions of each method to hold.

\begin{figure}
\includegraphics[width=0.9\columnwidth]{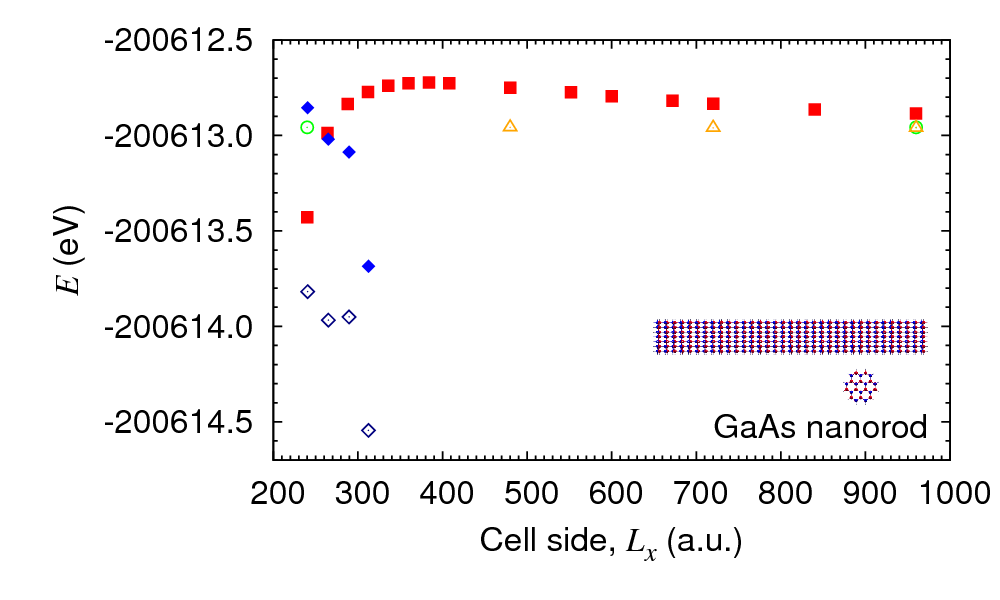}

\caption{\label{fig:nanorod-PBC_CC}Convergence of total energy with cell size
for GaAs Nanorod, showing results for Cutoff Coulomb (green circles),
Minimum Image Convention (orange triangles), Open Boundary Conditions
(blue diamonds, filled when smeared ions were used) and Periodic Boundary
Conditions (red squares) corrected using the Makov-Payne form (c)
fitted by least-squares fitting to $E_{0}$ and $B$. CC results are
independent of cell size for all $L_{x}$ greater than the rod length.
The MIC approach requires $L_{x}>2\times L_{\mathrm{rod}}$, so results
are only shown for $L_{x}\geq480a_{0}$.}

\end{figure}

The total energy of the the nanorod as a function of supercell side
length is shown in Figure~\ref{fig:nanorod-PBC_CC}. Here the default
supercell is not chosen to be cubic as this would be particularly
inefficient for such a high-aspect ratio nanostructure. We start with
$L_{x}=240a_{0}$, $L_{y},L_{z}=65a_{0}$ as the smallest cell able
to completely enclose the rod with around $10a_{0}$ padding in all
directions, and then $L_{y}$ and $L_{z}$ are scaled commensurately
with $L_{x}$. We see that in this case, with a highly polarisable
rod, the same Makov-Payne fit that successfully described the dipolar
systems in Section~\ref{sec:Convergence} now fails significantly
for all the cells studied here and returns an $E_{0}$ which does
not match the $L\to\infty$ limit, nor does it match the CC or MIC
results. The latter are well-converged with respect to $L_{x}$, and
are in good agreement with each other. However, again the OBC results
are strongly size-dependent, as a result of the approximations made
in order to obtain feasible computational effort at this large scale.
The validity of the convergence of the approximate methods starts
to break down beyond $L_{x}\sim300a_{0}$, resulting in significant
error.

\begin{figure}
\includegraphics[width=0.9\columnwidth]{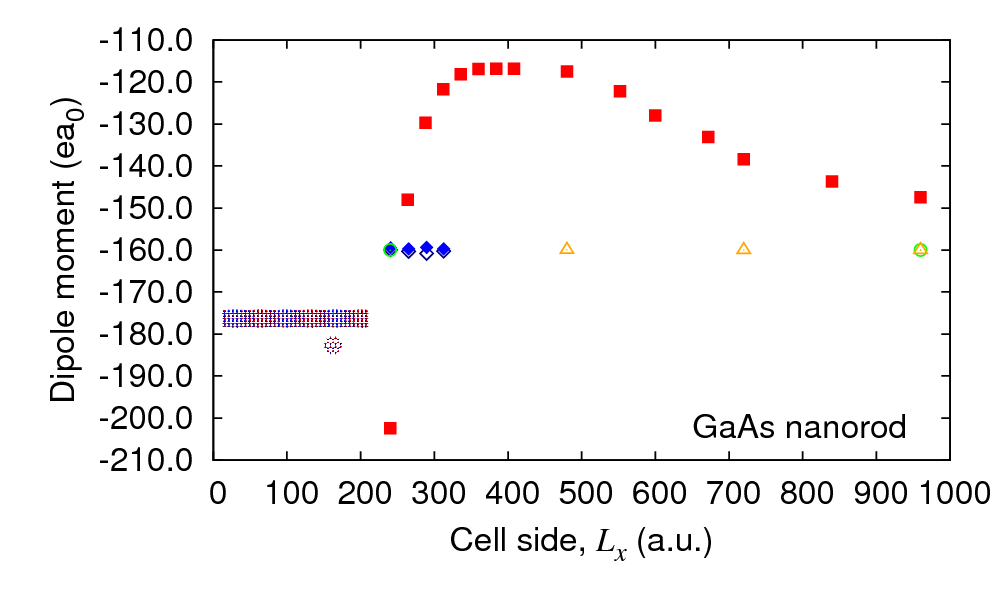}

\caption{\label{fig:nanorod-dipole}Dipole moment $d_{x}$ (see text) as a
function of cell size $L_{x}$ for a GaAs nanorod. Here the inset
illustration is shown approximately to-scale with the $L$-axis. The
exact form of $d_{x}(L)$ would depend on aspect ratio, and would
be difficult to accurately extrapolate to $L\to\infty$. The cutoff
Coulomb and MIC approaches obtain converged results for all cell sizes
large enough for their methods to hold, while the periodic results
converge only very slowly to this isolated value.}

\end{figure}

By examining the behaviour of the dipole moment of the rod along its
length, calculated as $d_{x}=\int_{{\cell}}x\, n(\mathbf{r})\,\mathrm{d}\mathbf{r}$,
we see immediately why this is the case: the dipole moment varies
strongly with cell size because of the induced polarisation caused
by the periodic images, as seen in Figure~\ref{fig:nanorod-dipole}.
The periodic images of the nanorods are all aligned, so if the rods
are very close end-to-end they will tend to increase the dipole moment.
However, if they are closer side-to-side the dipole field of the periodic
image (in the opposite direction to the polarisation, as viewed outside
the rod to its side) will tend to depolarise the rod and the dipole
moment will decrease. Therefore there is a strong dependence of $d_{x}$
on both $L_{x}$ and $L_{y},L_{z}$. Both of these are spurious effects
when one wishes to simulate an isolated rod. We see that all three
approaches, CC, MIC, and OBC, all correct these influences and obtain
the `correct' isolated result for $d_{x}$ even for small system sizes.

We therefore conclude that in such cases of large, polarisable systems
with a strong dipole moment, there is no choice but to use an approach
including the truncation of periodic images: in analogy to the study
of polar thin films \cite{bengtsson_dipole_1999}, a correction scheme
\emph{must} be employed if reliable results are to be obtained. We
have demonstrated that the approaches of Coulomb truncation and MIC
are suitable for this purpose. The inaccuracies inherent in the OBC
approach are particularly visible in the case of the nanorod, as the
very large box sizes cause the error associated with the evaluation
of the local pseudopotential to become unacceptably large. The error
due to the smeared ion representation is approximately $700\mu$eV/atom
and for the smallest box sizes it conveniently cancels most of the
error in the local pseudopotential, however for the larger simulation
cells the energy inevitably diverges. Although this divergence is
slow (compared to the total energy of the system), in the absence
of a monopole charge and the associated $O(L^{-1})$ term it makes
the OBC results unacceptably inaccurate for energies. Figure~\ref{fig:nanorod-dipole}
shows nevertheless that for other physical properties, such as the
dipole moment, it may be reliable.

\section{Conclusions\label{sec:Conclusion}}

We have described and applied three different methods, each with a
rather different theoretical basis, to the study of calculations on
charged and dipolar systems using linear-scaling density functional
theory under periodic boundary conditions. We have shown that with
each of these methods it is possible, while staying within a nominally
periodic formalism, to achieve the desired limit of equivalence of
any calculated properties to those of a single isolated system. 

In small systems, \emph{post-hoc} correction schemes are capable of
extrapolating to the isolated limit on the basis of several calculations
performed under PBCs, but only if simulation cells are used which
are very large compared to the system being studied. The first-order
term of the Makov-Payne correction, on its own, is inadequate for
accurate results, but by fitting the coefficient of the $O(L^{-3})$
term, one can acheive an accurate result for a cubic cell as long
as there is not a dipole moment present of comparable physical size
to the cell itself. This is clearly a highly computational expensive
approach due to the need to simulate several large cells to achieve
an accurate fit, and is not really practical for production calculations.
Fitting further coefficients of the Makov-Payne expansion tends to
reduce the accuracy by over-fitting to numerical noise.

However, we have also seen that the larger systems encountered in
large linear-scaling DFT calculations can behave very differently
from the small molecules in the test set. In particular, there is
scope in large systems for considerable long-range charge redistribution
\emph{in response to the effect of periodic images}, so reliable extrapolation
from simulations using a small unit cell are then impossible. In such
situations, one has no choice but to use a method that explicitly
negates the effect of periodic images.

Approaches which redefine the Coulomb potential to avoid periodic
interactions, either by using the Minimum Image Convention (whether
in the form proposed by Martyna and Tuckerman, or in the rather different
form by Genovese\emph{ \etal}) or the cutoff Coulomb method, have
been seen to rapidly converge to the isolated result as soon as the
conditions required as outlined above are met. In the case of the
MT formulation, this is that the size of the simulation cell be at
least twice the extent of the electron density in a given direction,
while in the Genovese form, this requirement is relaxed due to the
method being performed on what is effectively a padded real-space
grid.

The cutoff Coulomb approach is seen to produce accurately converged
results for a single-shot calculation, regardless of the size of the
simulation cell (as long as it is bigger than the extent of the nonzero
density). The only requirement is that the original cell must, for
the purposes of Fourier transforms, be embedded in a padded cell of
sufficient size. This generally entails quite a large temporary memory
requirement, and in small systems the performance of FFTs can become
the limiting factor on the speed of the whole calculation. However,
in large systems, where the Hartree calculation is generally not a
significant part of the total computational effort, this is no longer
an issue.

Finally, the use of Open Boundary Conditions, while exact in principle,
is seen to entail several further approximations in practice to render
it feasible. In particular, these enter into the evaluation of the
Dirichlet boundary values on the faces of the simulation cell, and
the use of a smeared-ion representation and the evaluation of the
local pseudopotential in real space. These approximations combine
to give a residual finite-size error notably larger than the other
methods can achieve. Furthermore, the multigrid approach to the Hartree
potential is computationally rather demanding and does not parallelise
as well as the rest of the approach. This makes the OBC approach the
least efficient method presented here. However, it has one major advantage
the others cannot match, namely that it can be used with an nonhomogeneous
dielectric constant, in the context of implicit solvent calculations.
For future calculations of this type, further investigation will be
required in order to develop means to calculate the boundary conditions
to higher accuracy -- such as by combining the multigrid OBC approach
with one of the other schemes here solely for the determination of
boundary conditions.

We noted also that two of the methods considered here can benefit
from similar speedups by suitable treatment of Fourier transforms
padded with zeroes. In both the cutoff Coulomb approach and the MIC
approach, there is a need to perform a FFT of the charge density to
reciprocal space under conditions where the value on most of the real-space
grid is known to be zero. In such cases, it has been shown that algorithms
can be designed which not only significantly reduce the computational
expense of such a transform but also reduce the memory usage by not
explicitly storing the zeros. The MIC implementation of Genovese\emph{
\etal} employs such an approach, but the Martyna-Tuckerman and cutoff
Coulomb approaches could in principle also do so. This would render
them all very similar in terms of computational cost, only marginally
above that of the original, uncorrected approach.
\begin{acknowledgments}
N.D.M.H and J.D. acknowledge the support of the Engineering and Physical
Sciences Research Council (EPSRC Grant Nos. EP/F010974/1, EP/G05567X/1
and EP/G055882/1) for postdoctoral funding through the HPC Software
Development call 2008/2009. P.D.H. and C.-K.S. acknowledge support
from the Royal Society in the form of University Research Fellowships.
The authors are grateful for the computing resources provided by Imperial
College's High Performance Computing service (CX2), and by Southampton's
iSolutions (Iridis3) which have enabled all the simulations presented
here. We would like to thank Stephen Fox for the structure of the
T4 lysozyme and Philip Avraam for the structure of the GaAs nanorod.
\end{acknowledgments}

\section*{Appendix A: Fourier Coefficients of the Cylindrically-Cutoff Interaction}

The integral for the Fourier components $\tilde{v}_{CC}(\mathbf{G})$
of the Coulomb interaction cut off over a cylinder of length $2L$
and radius $R$ can be written\begin{eqnarray*}
\tilde{v}_{CC}(\mathbf{G}) & = & \int_{cyl}\frac{e^{i\mathbf{G}.\mathbf{r}}}{r}\,\mathrm{d}\mathbf{r}\\
 & = & \int_{0}^{R}\int_{-L}^{L}\int_{0}^{2\pi}\frac{\rho\, e^{i(G_{\rho}\rho\sin\phi+G_{x}x)}}{\sqrt{\rho^{2}+x^{2}}}\mathrm{d}\phi\mathrm{d}x\mathrm{d}\rho\;.\end{eqnarray*}
Here we have taken the cylinder to be aligned along $x$, and taken
$G_{\rho}$ to lie in the $xy$-plane, without loss of generality.
To ensure that the resulting expression is finite and well-behaved
for all non-negative values of $\mathbf{G}$, we identify four regions
which must be treated separately: \begin{eqnarray*}
G_{\rho}>0 & , & G_{x}>0;\\
G_{\rho}=0 & , & G_{x}>0;\\
G_{\rho}>0 & , & G_{x}=0;\\
G_{\rho}=0 & , & G_{x}=0.\end{eqnarray*}
The latter three cases all allow significant simplification of the
integral and will be examined first. 

The $G_{\rho}=0,$ $G_{x}=0$ terms are the only ones where the integral
can be performed fully analytically:

\begin{eqnarray*}
\tilde{v}_{CC}(\mathbf{G}) & = & \int_{0}^{R}\int_{-L}^{L}\int_{0}^{2\pi}\frac{\rho}{\sqrt{\rho^{2}+x^{2}}}\,\mathrm{d}\phi\mathrm{d}x\mathrm{d}\rho\\
 & = & 4\pi\int_{\rho=0}^{R}\int_{x=0}^{L}\frac{\rho}{\sqrt{\rho^{2}+x^{2}}}\,\mathrm{d}x\mathrm{d}\rho\\
 & = & 4\pi\int_{0}^{L}\left(\sqrt{R^{2}+x^{2}}-x\right)\,\mathrm{d}x\\
 & = & 2\pi\Big[L(\sqrt{R^{2}+L^{2}}-L)+R^{2}\ln\left[\frac{L+\sqrt{R^{2}+L^{2}}}{R}\right]\Big]\end{eqnarray*}

The $G_{\rho}=0,$ $G_{x}>0$ terms can be rendered into a well-behaved
integral over $x$:

\begin{eqnarray*}
\tilde{v}_{CC}(\mathbf{G}) & = & \int_{0}^{R}\int_{-L}^{L}\int_{0}^{2\pi}\frac{\rho\, e^{iG_{x}x}}{\sqrt{\rho^{2}+x^{2}}}\,\mathrm{d}\phi\mathrm{d}x\mathrm{d}\rho\\
 & = & 4\pi\int_{0}^{R}\int_{0}^{L}\frac{\rho\,\cos G_{x}x}{\sqrt{\rho^{2}+x^{2}}}\,\mathrm{d}x\mathrm{d}\rho\\
 & = & 4\pi\int_{0}^{L}\left(\sqrt{R^{2}+x^{2}}-x\right)\cos G_{x}x\,\mathrm{d}x\end{eqnarray*}
which can be evaluated numerically with no significant difficulties

Similarly, the $G_{\rho}>0,$ $G_{x}=0$ terms can be made into a
well-behaved integral over $\rho$:

\begin{eqnarray*}
\tilde{v}_{CC}(\mathbf{G}) & = & \int_{0}^{R}\int_{-L}^{L}\int_{0}^{2\pi}\frac{\rho\, e^{iG_{\rho}\rho\sin\phi}}{\sqrt{\rho^{2}+x^{2}}}\,\mathrm{d}\phi\mathrm{d}x\mathrm{d}\rho\\
 & = & 2\int_{0}^{R}\int_{0}^{L}\int_{0}^{2\pi}\frac{\rho\cos[G_{\rho}\rho\sin\phi]}{\sqrt{\rho^{2}+x^{2}}}\,\mathrm{d}\phi\mathrm{d}x\mathrm{d}\rho\\
 & = & 4\pi\int_{0}^{R}\int_{0}^{L}\frac{\rho}{\sqrt{\rho^{2}+x^{2}}}J_{0}(G_{\rho}\rho)\,\mathrm{d}x\mathrm{d}\rho\\
 & = & 2\pi\int_{0}^{R}\ln\left[\frac{L+\sqrt{\rho^{2}+L^{2}}}{-L+\sqrt{\rho^{2}+L^{2}}}\right]\,\rho\, J_{0}(G_{\rho}\rho)\,\mathrm{d}\rho.\end{eqnarray*}
which also remains well-behaved over its range.

Finally, for $G_{\rho}>0,$ $G_{x}>0$, the integral cannot so easily
be put in a 1-dimensional form for easy evaluation. However, if the
cylinder length $L$ is first taken to infinity (effectively making
the interaction periodic in $x$), the integrals become tractable,
then the resulting answer can be convolved with a top-hat function
to retrieve the desired limits on the integral. The top hat function
is defined in terms of the Heaviside step function:\[
T(\mathbf{r})=\Theta(x+L)-\Theta(x-L)\;.\]
The transform of the Coulomb interaction for the infinite cylinder
would give\begin{eqnarray*}
\tilde{v}_{IC}(\mathbf{G}) & = & \int_{0}^{R}\int_{-\infty}^{\infty}\int_{0}^{2\pi}\frac{\rho\, e^{iG_{\rho}\rho\sin\phi+iG_{x}x}}{\sqrt{\rho^{2}+x^{2}}}\,\mathrm{d}\phi\mathrm{d}x\mathrm{d}\rho\;,\end{eqnarray*}
so we can write the transform of the finite cylinder as\begin{eqnarray*}
\tilde{v}_{CC}(\mathbf{G}) & = & \int_{0}^{R}\int_{-\infty}^{\infty}\int_{0}^{2\pi}T(\mathbf{r})\, v_{IC}(\mathbf{r})e^{i(G_{\rho}\rho\sin\phi+G_{x}x)}\,\mathrm{d}\phi\mathrm{d}x\rho\mathrm{d}\rho\;.\end{eqnarray*}
By the convolution theorem we can write the transform of the product
of two functions in real space as the convolution of these two functions
in reciprocal space. Using \textbf{$\mathbf{H}$} for our primed set
of reciprocal space coordinates we get: \[
\tilde{v}_{CC}(\mathbf{G})=\frac{1}{(2\pi)^{3}}\int\tilde{v}_{IC}(\mathbf{H})\tilde{T}(\mathbf{G}-\mathbf{H})\,\mathrm{d}^{3}\mathbf{H}\;.\]
All three integrals for $\tilde{v}_{IC}(\mathbf{H})$ can be done
analytically:

\begin{widetext}\begin{eqnarray*}
\tilde{v}_{IC}(\mathbf{H}) & = & \int_{0}^{R}\int_{-\infty}^{\infty}\int_{0}^{2\pi}\frac{\rho}{\sqrt{\rho^{2}+x^{2}}}\cos(H_{x}x)\cos(H_{\rho}\rho\sin\phi)\,\mathrm{d}\phi\mathrm{d}x\mathrm{d}\rho\\
 & = & 2\int_{0}^{R}\int_{0}^{2\pi}\rho\, K_{0}(H_{x}\rho)\cos(H_{\rho}\rho\sin\phi)\,\mathrm{d}\phi\mathrm{d}\rho\\
 & = & 4\pi\int_{0}^{R}\rho\, K_{0}(H_{x}\rho)J_{0}(H_{\rho}\rho)\,\mathrm{d}\rho\\
 & = & 4\pi\left[\frac{1+H_{\rho}R\, K_{0}(H_{x}R)\, J_{1}(H_{\rho}R)-H_{x}R\, K_{1}(H_{x}R)\, J_{0}(H_{\rho}R)}{H_{\rho}^{2}+H_{x}^{2}}\right]\end{eqnarray*}

\end{widetext}

This expression is in fact very simple to evaluate as it contains
no Bessel functions of higher order than 1. These can be rapidly evaluated
using accurate polynomial approximations over the domain required
for the integrals.

For the step function, the transform is well known\begin{eqnarray*}
\tilde{T}(\mathbf{G}) & = & \int_{-L}^{L}\exp[i\mathrm{G}_{x}x]\mathrm{d}x\,\delta(G_{\rho})\\
 & = & \frac{2\sin(G_{x}L)}{G_{x}}\delta(G_{\rho})\;.\end{eqnarray*}
Combining the two gives us \begin{eqnarray*}
\tilde{v}_{CC}(\mathbf{G}) & = & \frac{1}{(2\pi)^{3}}\int\frac{2\sin[(G_{x}-H_{x})L]}{G_{x}-H_{x}}\delta(G_{\rho}-H_{\rho})\tilde{v}_{IC}(\mathbf{H})\,\mathrm{d}^{3}\mathbf{H}\;.\end{eqnarray*}
After performing the $H_{\rho}$ integral to leave only $H_{\rho}=G_{\rho}$,
we obtain

\begin{widetext}\begin{eqnarray*}
\tilde{v}_{CC}(\mathbf{G}) & = & 4\int_{-\infty}^{\infty}\frac{\sin[(G_{x}-H_{x})L]}{(G_{x}-H_{x})}\\
 &  & \times\left[\frac{1+G_{\rho}R\, K_{0}(H_{x}R)\, J_{1}(G_{\rho}R)-H_{x}R\, K_{1}(H_{x}R)\, J_{0}(G_{\rho}R)}{G_{\rho}^{2}+H_{x}^{2}}\right]\mathrm{d}H_{x}\\
 & = & \frac{4\pi}{(G_{x}^{2}+G_{\rho}^{2})}\times\left(1-e^{-G_{\rho}L}(\frac{G_{x}}{G_{\rho}}\sin G_{x}L-\cos G_{x}L)\right)\\
 &  & +4\int_{-\infty}^{\infty}\frac{\sin[(G_{x}-H_{x})L][G_{\rho}R\, K_{0}(H_{x}R)\, J_{1}(G_{\rho}R)-H_{x}R\, K_{1}(H_{x}R)\, J_{0}(G_{\rho}R)]}{(G_{x}-H_{x})(G_{\rho}^{2}+H_{x}^{2})}\mathrm{d}H_{x}\end{eqnarray*}

\end{widetext}

Only the latter integral term needs to be calculated numerically.
One can see that as $R\rightarrow\infty$ and $L\rightarrow\infty$,
the modified Bessel function terms tend to zero, leaving only the
expected $4\pi/G^{2}$ behaviour from the first part. When performing
the integral numerically, the denominator damps out the oscillations
rapidly so the region of integration can be relatively small. A fairly
fine mesh must be used to capture the oscillations of the sinc function,
but not unmanageably so for the $G$-vectors typically required. We
used 200001 points in this work, ensuring convergence to 10 significant
figures for the largest $G_{x}$ and $L$ values required.

\section*{Appendix B: calculation of the local pseudopotential in real space}

The local pseudopotential $\vlocr$ can be evaluated in real space
as a sum of spherically-symmetrical contributions from all atomic
cores $I$, each located at $R_{I}$: \begin{equation}
\vlocr=\sum_{I}V_{\mathrm{locps},I}\left(\vert\bvec{r}-\bvec{R}_{I}\vert\right)\;.\end{equation}
To generate the local pseudopotential $\vlocIr$ due to core $I$
at a point $\bvec{r}$ in real space, the continuous Fourier transform
can be employed: \begin{eqnarray}
V_{\mathrm{locps},I}\left(\bvec{r}-\bvec{R}_{I}\right) & = & \frac{1}{{\left(2\pi\right)}^{3}}\int\vlocIg e^{i\bvec{G}\cdot\left(\bvec{r}-\bvec{R}_{I}\right)}d\bvec{G}\nonumber \\
 & = & \frac{1}{{\left(2\pi\right)}^{3}}\int\vlocIg e^{i\bvec{G}\cdot\bvec{x}}d\bvec{G}\;,\end{eqnarray}
where we have set $\bvec{x}=\bvec{r}-\bvec{R}_{I}$. We then use the
expansion of the plane wave $e^{i\bvec{G}\cdot\bvec{x}}$ in terms
of localised functions, to obtain: \begin{eqnarray*}
 &  & \vlocIx=\frac{1}{{\left(2\pi\right)}^{3}}\int\vlocIg\\
 & \times & \left[4\pi\sum_{l=0}^{\infty}\sum_{m=-l}^{l}i^{l}\jmath_{l}\left(Gx\right)Z_{lm}\left(\Omega_{\bvec{G}}\right)Z_{lm}\left(\Omega_{\bvec{x}}\right)\right]d\bvec{G}\;,\end{eqnarray*}
where $\jmath_{l}$ are spherical Bessel functions of the first kind
and $Z_{lm}$ are the real spherical harmonics. A simple rearrangement
leads to \begin{eqnarray}
\vlocIx & = & \frac{4\pi}{{\left(2\pi\right)}^{3}}\sum_{l=0}^{\infty}\sum_{m=-l}^{l}i^{l}Z_{lm}\left(\Omega_{\bvec{x}}\right)\nonumber \\
 & \times & {\int\vlocIg\jmath_{l}\left(Gx\right)Z_{lm}\left(\Omega_{\bvec{G}}\right)d\bvec{G}}\;.\label{eq:localpseudo}\end{eqnarray}
After changing into spherical polar coordinates and applying the orthonormality
property of spherical harmonics, the above expression simplifies to
a spherically-symmetric form: \begin{equation}
\vlocIxscalar=\frac{4\pi}{{\left(2\pi\right)}^{3}}\int\limits _{0}^{\infty}\vlocIgscalar{}\frac{\sin\left(Gx\right)}{x}G\, dG\;.\label{eq:vloci}\end{equation}
In practice, it is sufficient to evaluate this expression once, for
every ionic species $s(I)$, rather than for every core $I$, on a
fine radial grid with $x$ ranging from 0 to a maximum value dictated
by the size of the simulation cell in use. A finite upper limit, $\gcut$,
corresponding to the longest vector representable on the reciprocal
grid, should be used in the integral in Eq~(\ref{eq:vloci}), in
order to avoid aliasing when transforming from reciprocal to real
space.

The numerical evaluation of the integral in Eq.~(\ref{eq:vloci})
is not straightforward. One source of difficulties is the oscillatory
nature of $\sin\left(Gx\right)$. For larger cells, the oscillations
become so rapid that the resolution with which the reciprocal-space
coefficients $\vlocsgscalar$ of the pseudopotential are provided,
typically 0.05 \AA{}$^{-1}$, is not sufficient and it becomes necessary
to interpolate $\vlocsgscalar$, and the whole integrand, in between
these points. Another difficulty is caused by the singularity in $\vlocsgscalar$
as $G\to0$, where the behaviour of $\vlocsgscalar$ approaches that
of $-Z_{s}/G^{2}$ (where $Z_{s}$ is the charge of the core of species
$s$). Although the integral is convergent, this singularity cannot
be numerically integrated in an accurate fashion, and it also contributes
to making the abovementioned interpolation inaccurate at low $G$'s.
This is partially alleviated by subtracting the Coulombic potential,
$-Z_{s}/G^{2}$, before interpolating to the fine radial reciprocal-space
grid, and then adding it back but the residual numerical inaccuracy
leads to a near-constant shift of the obtained real-space pseudopotential,
which in turn results in errors in the total energy in the order 0.01\%.

To address this problem the pseudopotential can be partitioned into
a short-range and a long-range term, similarly as in Eq.~(\ref{eq:partition}).
This leads to \begin{eqnarray}
\vlocIxscalar & = & -\frac{Z_{s}\operatorname{erf}{\left(\alpha{}x\right)}}{x}+\frac{4\pi}{{\left(2\pi\right)}^{3}}\int\limits _{0}^{\infty}\vlocIgscalar{}\nonumber \\
 & \times & \left[1-\exp{\left(\frac{-G^{2}}{4\alpha^{2}}\right)}\right]\frac{\sin\left(Gx\right)}{x}G\, dG\;,\label{eq:vlocisrlr}\end{eqnarray}
where the first term, with the error function, is the long range part
and the second is the short range part and $\alpha$ is an adjustable
parameter that controls where the transition between short-range and
long-range takes place.

Owing to the $\left[1-\exp{\left(\frac{-G^{2}}{4\alpha^{2}}\right)}\right]$
factor, the singularity at $G=0$ is avoided in the same way as in
Eq.~(\ref{eq:short-zero}) and the integral can be accurately evaluated
numerically, provided $\alpha$ is large enough. Smaller values of
$\alpha$ make the numerical integration less accurate, because the
oscillations at low values of $G$ increase in magnitude. Larger values
of $\alpha$ increase the accuracy of the integration, however, they
lead to a faster decay of the reciprocal-space term and cause the
long-range behaviour to be increasingly more dictated by the first
term in the RHS of Eq.~(\ref{eq:vlocisrlr}). As this term is calculated
in real space, it lacks the oscillations that are expected to be present
in the pseudpotential at large $x$, due to the finite value of $\gcut$,
causing aliasing. For this reason $\alpha$ needs to be as small as
possible, without negatively impacting on the accuracy of the numerical
integration.

The accuracy of the approach can be assessed by comparing the real-space
tail of the obtained pseudopotential with the Coulombic potential.
Since the obtained pseudopotential is expected to oscillate slightly
so that it takes values above and below $-Z_{s}/x$, a good measure
of accuracy, which we will call $b$, is $\left<\dfrac{\vlocsxscalar-(-Z_{s}/x)}{-Z_{s}/x}\right>$,
where the average runs over the real-space tail of the pseudopotential,
from, say, 5\,$a_{0}$ to the maximum $x$ for which $\vlocsxscalar$
is evaluated. Ideally, $b$ should be zero. Numerical inaccuracy will
cause a shift in $\vlocsxscalar$ which will present itself as a finite,
non-zero value of $b$. Direct numerical integration of Eq.~(\ref{eq:vloci})
using various high order quadrature schemes results in values of $b$
in the order of 0.01, which can be reduced by an order of magnitude
by interpolating to a very fine radial reciprocal-space grid. Subtracting
the Coulombic potential and integrating only the difference between
$\vlocsgscalar$ and the Coulombic potential numerically, while analytically
integrating the remaining part reduces $b$ to about 0.0005. Application
of the proposed approach Eq.~(\ref{eq:vlocisrlr}) yields $b=5\times10^{-8}$
for $\alpha=0.5/l$ and $b=3\times10^{-9}$ for $\alpha=0.1/l$, where
$l$ is the length of the simulation cell. The total energy is then
insensitive (to more than 0.0001\%) to the choice of $\alpha$, provided
it is in a wide {}``reasonable'' range of $0.1/l-2/l$.

\bibliographystyle{aipnum4-1} \bibliographystyle{aipnum4-1}

\end{document}